\documentclass[]{elsarticle}
\usepackage{graphicx} 
\usepackage{a4wide}
\usepackage{csquotes}

\usepackage[dvipsnames]{xcolor}
\usepackage{graphicx,color,booktabs,listings,caption,subcaption,tikz,amsmath,amssymb,placeins,subfiles,tabto, parskip}
\usepackage[inkscapelatex=false]{svg}

\usepackage[titletoc]{appendix}
\usepackage{algpseudocode}
\usepackage{algorithm}
\usepackage{esvect}
\usepackage{gensymb}
\usepackage{inconsolata}
\usepackage{hyperref}
\usepackage{natbib}
\hypersetup{
    colorlinks=true,
    linkcolor=blue,
    filecolor=magenta,      
    urlcolor=cyan,
    pdftitle={Overleaf Example},
    pdfpagemode=FullScreen,
    }
\usepackage{pdfpages}

\usepackage{arydshln}
\newcommand{\D}{\mathrm{d}}

\DeclareMathOperator*{\argmin}{arg\,min}

\title{Reduced Subgrid Scale Terms in Three-Dimensional Turbulence}
\date{\today}

\begin{document}
\begin{frontmatter}
\author[inst1]{Rik Hoekstra \corref{cor1}}
\affiliation[inst1]{organization={Centrum Wiskunde \& Informatica, Scientific Computing group},
            addressline={Science Park 123}, 
            city={Amsterdam},
            postcode={1098 XG}, 
            country={the Netherlands}}

\author[inst1]{Wouter Edeling}
\cortext[cor1]{Corresponding author, e-mail address: rik.hoekstra@cwi.nl}

\begin{abstract} 
Large eddy simulation (LES) has become a central technique for simulating turbulent flows in engineering and applied sciences, offering a compromise between accuracy and computational cost by resolving large scale motions and modeling the effects of smaller, unresolved scales through a subgrid scale (SGS) model. The fidelity and robustness of LES depend critically on the SGS model, particularly in coarse simulations where much of the turbulence spectrum remains unresolved.

In this work, we extend the tau-orthogonal (TO) method, a data-driven SGS modeling framework, to three-dimensional turbulent flows. The method reformulates the high-dimensional SGS closure problem as a low-dimensional prediction task focused on scale-aware quantities of interest (QoIs). We extend the model to incorporate QoI-state dependence and temporal correlations by combining regularized least-squares regression with a multivariate Gaussian residual model. This yields a simple yet effective stochastic time-series prediction model (the LRS model), with orders-of-magnitude fewer parameters than typical deep learning approaches which try to directly learn the high-dimensional SGS closure.

We demonstrate the effectiveness of the TO LRS model in three-dimensional forced isotropic turbulence and turbulent channel flow. The model achieves accurate long-term QoI distributions, robust performance across hyperparameter settings, and good reproduction of key flow features such as kinetic energy spectra and coherent structures, despite being trained solely on QoI trajectories. Comparisons against classical SGS models, including Smagorinsky and WALE formulations, highlight the TO LRS model’s balance of accuracy and computational efficiency.
\end{abstract}

\begin{keyword}
reduced subgrid scale modelling \sep large eddy simulation \sep turbulence \sep stochastic
\end{keyword}

\end{frontmatter}

\section{Introduction}
Turbulent flows are everywhere. They play an important role in mixing of the Earth's atmosphere and oceans, as well as in aerodynamics. These flows span a vast range of spatial and temporal scales, which are hard to fully resolve in real-world applications. Capturing the smallest scales would require a number of grid points far beyond current computing capabilities. To circumvent this problem in turbulence simulations, large eddy simulation (LES) has become a popular method \cite{lesieur_large-eddy_2005}. In LES, only the large-scale motions are explicitly resolved; whereas the smaller, unresolved scales are modeled using a subgrid scale (SGS) model. The performance of these large eddy simulations, depends heavily on the quality of the SGS model and the portion of the turbulent scales that needs to be modeled.

Classical SGS models, such as the Smagorinsky model \cite{smagorinsky_general_1963} and its dynamic variants \cite{germano1991dynamic} have been widely employed due to their simplicity and empirical success, especially when the majority of the turbulent scales are resolved. They model the influence of the subgrid scales as a function of the resolved field. While these models are useful, they have notable limitations when a larger portion of the scales needs to be modeled. In particular, they often struggle to adapt to different flow conditions, and they tend to oversimplify turbulent dynamics.

At the other end of the modeling spectrum lie Reynolds averaged Navier–Stokes (RANS) models, which model all turbulent fluctuations. The RANS equations essentially solve none of the turbulent scales, but are computationally cheap. A wide variety of RANS models have been developed to handle different flow scenarios \cite{girimaji_turbulence_2024}. However, they cannot easily leverage the growing availability of computational power, which now makes it feasible to resolve a substantial portion of the turbulence spectrum in many applications.

This evolving computational landscape opens the door for more flexible and expressive LES-SGS models, that can faithfully represent subgrid dynamics regardless of how much of the turbulence spectrum is resolved. This motivates our focus on test cases designed to emulate very coarse LES conditions. 

Recent advances in data-driven modeling have offered promising tools to construct flexible SGS models for LES, particularly through deep learning techniques \cite{gimenez_multiscale_2025, li_fourier_2022, saura_subgrid_2022, kurz_harnessing_2025, liu_investigation_2022, park_toward_2021}. These models learn subgrid dynamics from training data sets, which can be obtained from high-resolution simulations. While these models provide greater adaptability to complex turbulent behaviors compared to classical SGS models, they also come with drawbacks. They are computationally more expensive and often act as ``black boxes", lacking interpretability and guarantees on generalization. Therefore, addressing these challenges is crucial for the wider adoption of deep learning in SGS modeling \cite{sanderse_scientific_2024}.

Over the recent years, several approaches have been explored to mitigate the drawbacks of deep learned SGS models. We mention some of these approaches to sketch the background for the methods introduced in this paper. Major efforts have been made on constraining the SGS models with known physical laws, using soft constraints or hard constraints \cite{guan_learning_2023, van_gastelen_energy-conserving_2023} which do considerably improve the generalization of the models. Furthermore, it has turned out to be of importance that the training data, used to train the SGS model, is as consistent as possible with the data the model needs to process when employed in a solver.
\cite{kurz_Deep_reinforcement_2023} and \cite{agdestein_discretize_2024} have addressed the training-model data inconsistency in the filters of implicit LES, and explicit LES respectively. 

Besides that, the community has gained insights into the advantages of \begin{it}a-posteriori\end{it} learning compared to \begin{it}a-priori learning\end{it}, highlighting a trade-off between training costs and model accuracy \cite{list2022learned, list_differentiability_2025}. In a-priori learning, the SGS model is trained to reproduce subgrid terms given resolved flow snapshots, without feedback from the full simulation; whereas in a-posteriori learning, the model is trained in the context of a running solver, a computationally very costly procedure. A particularly promising middle ground between these strategies could involve implementing a nudging procedure during the learning process \cite{rasp_coupled_2020, ling_numerically_2025}.

Additionally, slightly changing the framework to “ideal LES” \cite{langford1999IDEALLES} leads naturally to the concept of stochastic SGS models. This framework emphasizes that, given the resolved scales, the unresolved scales might still be in any configuration. Therefore, predicting the next state of the system should not be seen as a deterministic mapping. This perspective encourages the search for stochastic closure terms, which have been developed in simple form by \cite{guillaumin_stochastic-deep_2021} or can be obtained by training generative deep networks \cite{perezhogin_generative_2023}. The ``ideal" LES framework has also been unified with the field of data assimilation to obtain a simple stochastic SGS model \cite{ephrati2024probabilistic}.

Our research primarily addresses the high computational costs associated with training and evaluating the large deep learning models. Our approach, the tau-orthogonal method, is grounded in the observation that many practical applications of LES focus on a limited set of quantities of interest (QoIs)—such as average energy or enstrophy—rather than the full, high-dimensional flow field. By focusing on these QoIs, we reformulated the SGS modeling task as a low-dimensional learning problem, significantly reducing the computational complexity while improving model interpretability \cite{edeling_reducing_2020}.

The key to our approach lies in representing unresolved dynamics using a minimal set of scalar time series, one for each QoI. To obtain statistics of these time series, we use a nudging approach that forces LES simulations toward reference QoI trajectories. In earlier work \cite{hoekstra2024Reduced_data-driven}, we demonstrated that this method can successfully reproduce the long-term distributions of four QoIs in a simple two-dimensional turbulence test case, using a random noise model for the time series, provided sufficient training data is available.

In this paper, we derive stochastic reduced SGS models for three-dimensional turbulent flows. Besides increasing the problem dimension, we move to a staggered-grid solver with a new coarse-graining procedure, demonstrating that Fourier-space-based, scale-aware QoIs can be coupled to solvers operating entirely in physical space.
To better capture the temporal correlations of QoIs in three dimensions, we introduce a linear regression model, combining a small number of lagged QoI values with an additive noise term. This keeps the learning task low-dimensional and tractable, independent of spatial resolution. In contrast, applying conventional deep learning approaches to map directly from the resolved field to the SGS term becomes even less practical in three dimensions, as both input and output dimensions grow cubically with grid size, requiring vast datasets and significant compute resources.

The remainder of this paper is organized as follows. Section~\ref{section2: governing equations} introduces the coarse-grained Navier–Stokes equations and the QoIs. Section~\ref{section: TO method} presents our methodology in detail. Sections~\ref{section 4: HIT} and~\ref{section 5: Channel} validate the proposed models on three-dimensional isotropic turbulence and channel flow test cases, comparing their performance to classical SGS models. We conclude in Section~\ref{section 6: conclusion}.

\section{Governing Equations} \label{section2: governing equations}

The incompressible Navier-Stokes equations describe the time evolution of the velocity field $\mathbf{u} = (u_1, u_2, u_3)^T$ and the pressure field $p$ in three dimensions. They are given by:

\begin{equation}
    \nabla \cdot \mathbf{u} = 0,
\end{equation}
\begin{equation}
    \frac{\partial \mathbf{u}}{\partial t} + \nabla \cdot (\mathbf{u} \mathbf{u}^T) = -\nabla p + \nu \nabla^2 \mathbf{u} + \mathbf{f},
\end{equation}
where $\nu$ denotes the kinematic viscosity, and $\mathbf{f}$ is a forcing term used to sustain turbulence.

We employ an energy-conserving, incompressible Navier-Stokes solver \cite{agdesteinincompressiblenavierstokesjl_2024} to solve these equations. This solver discretizes the equations on a staggered Cartesian grid.
On this grid, velocity components are defined at cell faces, whereas pressure is defined at cell centers.
A projection method eliminates the pressure term, resulting in the discretized pressure-free evolution equation (for details, see \cite{agdestein_discretize_2024}):

\begin{align}
    \frac{d\mathbf{u}}{dt} &= P \mathbf{F}(\mathbf{u}), \\
    \mathbf{F}(\mathbf{u})&=-\mathbf{G} (\mathbf{u} \mathbf{u}^T) + \nu \mathbf{D} \mathbf{u} + \mathbf{f},
\end{align}
where \( P \) is the pressure projection operator, $\mathbf{D}$ is the discretized divergence operator, $\mathbf{G}$ is the discretized gradient operator, and $\mathbf{F}(\mathbf{u})$ represents the combined effects of the convective, viscous, and forcing terms. The pressure projection operator ensures that the velocity field $\mathbf{u}$ remains divergence-free. 

Direct numerical simulations (DNS) of simple turbulent flows are achievable using this solver. However, resolving all scales of motion requires a very fine grid resolution, rendering DNS computationally demanding.

\begin{figure}
    \centering
    \begin{minipage}{.4\textwidth}
\begin{tikzpicture}
\draw (0,0) grid (4,4);
\draw[shift={(-.5,-.5)}, double, double distance=2.2pt, line cap=round, dash pattern=on 0pt off 1cm] (1,1) grid (4,4);
\foreach \x in {1,...,4}
    \foreach \y in {1,...,4}
        {
        \ifthenelse{\x = 2 \and \y <3}{\draw [->, teal] (\x-0.1, \y-0.5) to ++(0.2, 0);}
        {\draw [->, black] (\x-0.1, \y-0.5) to ++(0.2, 0);}
        \ifthenelse{\y < 3}{\draw [->, black] (\x-0.5, \y-0.1) to ++(0.0, 0.2);}
        {\draw [->, black] (\x-0.5, \y-0.1) to ++(0.0, 0.2);}
        }
\draw [->, white] (4, 4) to ++(0.2, 0.2);

\end{tikzpicture}
\end{minipage}
\begin{minipage}{.4\textwidth}
\begin{tikzpicture}
\begin{scope}[scale = 2]
\draw (0,0) grid (2,2);
\draw[shift={(-.5,-.5)}, double, double distance=2.2pt, line cap=round, dash pattern=on 0pt off 1.99cm] (1,1) grid (2,2);
\foreach \x in {1,...,2}
    \foreach \y in {1,...,2}
        { \ifthenelse{\x = 1 \and \y=1}
        {\draw [->, teal] (\x-0.1, \y-0.5) to ++(0.2, 0);}
        {\draw [->] (\x-0.1, \y-0.5) to ++(0.2, 0);}
        \draw [->] (\x-0.5, \y-0.1) to ++(0.0, 0.2);
        }
\end{scope}
\end{tikzpicture}
\end{minipage}
    \caption{Coarse-graining a two-dimensional staggered grid with face-averaging. The arrows indicate velocity components, and the dot denotes the location of the pressure.}
    \label{fig: face-averaging on staggered grid}
\end{figure}
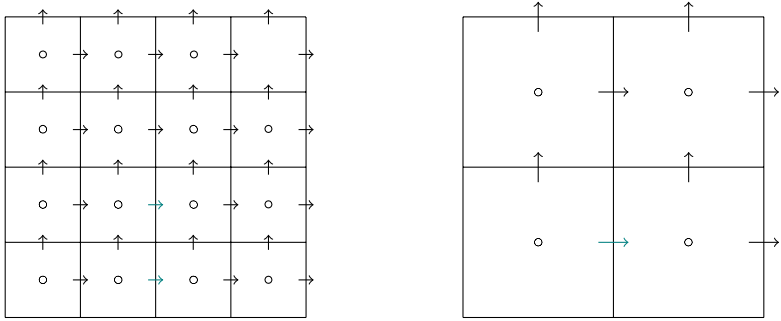

\subsection{Coarse-graining and SGS term}
To alleviate the computational cost of the simulations, we aim to solve the equations on a coarser grid. In order to move from the solution at DNS resolution to a solution on a coarser grid we used a filtering operation, denoted by
\begin{equation}
    \bar{\mathbf{u}} = \Phi \mathbf{u},
\end{equation}
where $\mathbf{u}$ represents the DNS velocity field, and $\bar{\mathbf{u}}$ is the coarse-grained velocity field. Following \cite{agdestein_discretize_2024} we used a face-averaging filter, which averages velocity components on the faces of coarser cells, as illustrated in Figure \ref{fig: face-averaging on staggered grid}. This filter maintains the divergence-free property of the coarse-grained velocity field, in contrast to standard volume averaging-filters.

Applying this filtering operation to the pressure-free evolution equation yields the filtered equation:
\begin{equation}
    \frac{d\bar{\mathbf{u}}}{dt} = \bar{P} \bar{\mathbf{F}}(\bar{\mathbf{u}}) + \mathbf{c}(\mathbf{u}, \bar{\mathbf{u}}),
\end{equation}
where $\bar{P}$ and $\bar{\boldsymbol{F}}$ are operators on the coarse grid, and $\mathbf{c}(\mathbf{u}, \bar{\mathbf{u}}) = \Phi P \mathbf{F}(\mathbf{u}) - \bar{P} \bar{\mathbf{F}}(\bar{\mathbf{u}})$ is the commutator error arising from the filtering operation. This term, commonly referred to as the subgrid scale (SGS) term, effectively represents the influence of unresolved small scales on the resolved scales. Notably, the SGS term is also divergence-free.

However, the SGS term depends on the DNS field, which is not available in coarse simulations. Consequently, it must be approximated by a model $\mathbf{m}$ that depends solely on resolved scales. This leads to the low-fidelity model equation:
\begin{equation} \label{filterd equations}
    \frac{d\mathbf{v}}{dt} = \bar{P} (\bar{\mathbf{F}}(\mathbf{v}) + \mathbf{m}(\mathbf{v})),
\end{equation}
where $\mathbf{v}$ denotes the solution of the low-fidelity simulation. We include the SGS model within the projection to ensure $\mathbf{v}$ remains divergence-free.

The main challenge of LES lies in developing an SGS model and solver that satisfy the approximation:
\begin{equation}
    \mathbf{v}(t) \approx \bar{\mathbf{u}}(t).
\end{equation}
The interpretation of this approximation varies. While some models aim to replicate the filtered DNS solution pointwise, a task rendered impractical over extended simulations due to the chaotic nature of turbulence, others focus on reproducing statistical properties. In this study, we focus on the time evolution of integrated quantities of the solution fields, which will be elaborated in the subsequent section.

\subsection{Quantities of interest and scale-awareness}

Instead of reproducing the full flow field, we focus on a limited set of physically relevant quantities of interest (QoIs) that encapsulate the key turbulent dynamics. We select QoIs, such as kinetic energy and enstrophy, for their ability to characterize turbulence across different scales. By targeting these integrated quantities, our approach reduces the dimensionality of the closure problem by several orders of magnitude, irrespective of the number of spatial dimensions in the problem, or the degrees of freedom in the filtered governing equations.

A QoI is defined as an integrated function of the coarse-grained solution:

\begin{equation} \label{eq: qoi def}
    Q = \int_{\Omega} q(\mathbf{v}) \D \boldsymbol{x}.
\end{equation}
In this paper we examined two well-known quantities of such form; kinetic energy and enstrophy.\\
The total kinetic energy within the computational domain $\Omega$ is given by

\begin{equation}
    E = \frac{1}{2} \int_\Omega \|\bar{\mathbf{u}}\|^2 \, \D \boldsymbol{x}.
\end{equation}

To capture dynamics across scales, we decomposed the energy into wavenumber bins in Fourier space. To this end, we introduce the total energy in Fourier space as 
\begin{equation}
    \hat{E}
    =  \frac{1}{2} \frac{\lvert \Omega \rvert}{(N_xN_yN_z)^2} \sum_{\mathbf{k}} \hat{\mathbf{u}}_\mathbf{k} \cdot \text{conj}(\hat{\mathbf{u}}_\mathbf{k}),
\end{equation}
where $\lvert \Omega \rvert$ denotes the volume of the computational domain, $\hat{\mathbf{u}}$ is the Fourier-transformed velocity field, $\mathbf{k}$ are the wavenumber vectors in Fourier space, and $N_x$ denotes the number of grid points in the $x$-direction used for the Fourier expansion. This expression follows from applying quadrature in Fourier space, see \ref{app: Fourier quadature}. We define the energy in wavenumber bin $[l, m]$ as follows

\begin{equation} \label{eq: scale aware energy}
    \hat{E}[l, m]
    =  \frac{1}{2} \frac{\lvert \Omega \rvert}{(N_xN_yN_z)^2} \sum_{\mathbf{k}} \hat{R}_{[l, m]}\hat{\mathbf{u}}_\mathbf{k} \cdot \text{conj}(\hat{R}_{[l, m]}\hat{\mathbf{u}}_\mathbf{k}).
\end{equation}

Here we introduced the scale-aware sharp Fourier filter $R_{[l,m]}$
\begin{align}
    R_{[l,m]} &= \mathcal{F}^{-1} \hat{R}_{[l,m]} \mathcal{F}, \nonumber\\
    \hat{R}_{[l, m]} \,\hat{\mathbf{u}}_{\bf k} &=
    \begin{cases}
        \hat{\mathbf{u}}_{\mathbf{k}} & \mathrm{if} \quad l-\frac{1}{2} \leq \lVert{\bf k}\rVert_2 < m + \frac{1}{2} \\
        0 & \mathrm{otherwise}
    \end{cases}.
\end{align}

Similarly, the enstrophy is defined as the integral over the squared vorticity

\begin{equation}
    Z = \int_\Omega \|\bar{\boldsymbol{\omega}}\|^2 \, \D\boldsymbol{x} = \int_\Omega \|\nabla \times \bar{\mathbf{u}}\|^2 \, \D\boldsymbol{x},
\end{equation}

which has a corresponding scale-aware form:

\begin{equation}
    \hat{Z}[l, m] = \frac{\lvert \Omega \rvert}{(N_xN_yN_z)^2} \sum_{\mathbf{k}} \hat{R}_{[l, m]}\hat{\boldsymbol{\omega}}_\mathbf{k} \cdot \text{conj}(\hat{R}_{[l, m]}\hat{\boldsymbol{\omega}}_\mathbf{k}). \label{eq: scale aware enstrophy}
\end{equation}

where $\hat{\boldsymbol{\omega}}$ is the Fourier-transformed vorticity. This spectral decomposition of energy and enstrophy allows us to analyze turbulence characteristics across distinct spatial scales, improving the accuracy of the SGS model by focusing on key statistical properties of turbulent flow.

\section{The tau-orthogonal method for reduced SGS modeling} \label{section: TO method}

The tau-orthogonal (TO) method offers a computationally efficient approach to SGS modeling focusing on a limited set of QoIs. Instead of modeling the SGS term pointwise, the TO method approximates the SGS term as a weighted sum of spatial patterns, each corresponding to a QoI:

\begin{equation} \label{eq: TO anzats}
    \mathbf{m}(\mathbf{v}, t) = \sum_{i=1}^{N_Q} \tau_i(t) \mathbf{O}_i(\mathbf{v}),
\end{equation}

where $N_Q$ is the number of QoIs, $\tau_i(t)$ are time-dependent scalar coefficients representing the unresolved contributions to the $i$-th QoI, and $\mathbf{O}_i(\mathbf{v})$ are spatial patterns chosen such that each QoI is independently controlled. 

We derive the TO method for a periodic computational domain. By projecting the governing equations onto the selected QoIs, we derive a reduced system that evolves their dynamics directly. We start from the ODEs for the selected QoIs:

\begin{equation}
    \frac{\D Q_i}{\D t} = 
    \int_\Omega \frac{{\D q_i(\mathbf{v})}}{\D t} \D \boldsymbol{x}
    =
    \int_\Omega \mathbf{V}_i \cdot \frac{{\partial R_i \mathbf{v}}}{\partial t} \D \boldsymbol{x}.
\end{equation}
where $R_i$ is the scale-aware filter used for the $i$-th QoI, and $\mathbf{V}_i$ is the weak derivative of $q_i$ w.r.t. $R_i \mathbf{v}$, 
i.e.\ $\int_\Omega q_i \frac{\partial \boldsymbol{\varphi}}{\partial R_i \mathbf{v}} \D \boldsymbol{x}=-\int_\Omega \mathbf{V}_i \boldsymbol{\varphi} \D \boldsymbol{x}$ for the set of test functions $\boldsymbol{\varphi}$.

We substitute the filtered equation \eqref{filterd equations} as expression for the time derivative: 
\begin{equation}
    \frac{\mathrm{d} Q_i}{\mathrm{d} t} = \int_\Omega \mathbf{V}_i \cdot R_i \bar{P} (\bar{\mathbf{F}}(\mathbf{v}) + \mathbf{m}(\mathbf{v})) \D \boldsymbol{x}.
\end{equation}

We construct the spatial patterns in \eqref{eq: TO anzats} such that they each affect only a single QoI. This is achieved by enforcing the following orthogonality conditions:

\begin{equation} \label{eq: orthogonality}
    \int_\Omega \mathbf{V}_i \cdot \mathbf{O}_j \, d\boldsymbol{x} = 0 \quad \text{for } i \neq j.
\end{equation}

Decomposing the evolution of each QoI into resolved and unresolved contributions leads to a system, where only the $\tau_i$ time series are unclosed:

\begin{equation} \label{eq: closed ODEs}
    \frac{\mathrm{d} Q_i}{\mathrm{d} t} =  \frac{\mathrm{d} Q_i^r}{\mathrm{d} t} +  \frac{\mathrm{d} Q_i^u}{\mathrm{d} t} 
    =
    \frac{\mathrm{d} Q_i^r}{\mathrm{d} t} +  \tau_i(t)
     \int_\Omega \bar{P} \mathbf{V}_i \cdot R_i \mathbf{O}_i \D \boldsymbol{x}.
\end{equation}

This formulation allows us to model unresolved dynamics as a low-dimensional time-series problem, shifting the burden of subgrid modeling from high-dimensional spatial fields to a set of reduced-order coefficients.

\subsection{Constructing spatial patterns}

To fulfill the orthogonality constraints \eqref{eq: orthogonality}, we construct the basis functions $\mathbf{O}_i$ using weak derivatives $\mathbf{V}_i$ of the QoIs. For the scale-aware energy we derive, using \eqref{eq: scale aware energy}:
\begin{align}
    \frac{\D E_{[l,m]}}{\D t} &= \frac{1}{2}C\int_\Omega \sum_{\alpha=1}^3 \frac{\partial (R_{[l,m]} {v}_\alpha  R_{[l,m]} {v}_\alpha)}{\partial R_{[l,m]} {v}_\alpha}  \frac{\partial R_{[l,m]} {v}_\alpha}{\partial t} \D \boldsymbol{x} \nonumber\\
    &= C\int_\Omega R_{[l,m]} \mathbf{v} \cdot \frac{\partial R_{[l,m]} \mathbf{v}}{\partial t} \D \boldsymbol{x},
\end{align}
where $C=\lvert \Omega \rvert/(N_xN_yN_z)^2$. Therefore $\mathbf{V}_i = R_{[l,m]} \mathbf{v}$.

Finding the weak derivative for the scale-aware enstrophy is less straightforward. Using the chain rule and integration by parts we find:
\begin{align}
    \frac{\D Z_{[l,m]}}{\D t} &= C \int_\Omega \sum_{\alpha=1}^3 \frac{\partial (R_{[l,m]} {\omega}_\alpha  R_{[l,m]} {\omega}_\alpha)}{\partial R_{[l,m]} {\omega}_\alpha}  \frac{\partial R_{[l,m]} {\omega}_\alpha}{\partial t} \D \boldsymbol{x} \nonumber\\
    &= C \int_\Omega 2 R_{[l,m]} \boldsymbol{\omega} \cdot  \frac{\partial R_{[l,m]} ( \nabla \times  \mathbf{v})}{\partial t} \D \boldsymbol{x} \nonumber\\ 
    &= C \int_\Omega (2 \nabla \times R_{[l,m]}\boldsymbol{\omega}) \cdot \frac{\partial R_{[l,m]} \mathbf{v}}{\partial t} \D \boldsymbol{x},
\end{align}
where $C=\lvert \Omega \rvert/(N_xN_yN_z)^2$.
So for the enstrophy $\mathbf{V}_i = 2 \nabla \times R_{[l,m]}\boldsymbol{\omega}$.

Finally, the spatial patterns are constructed as:
\begin{equation}
    \mathbf{O}_i = \sum_{j=1}^{N_Q} c_{i,j}T_j(\mathbf{v}, \mathbf{x}), \quad \textrm{for}\;\;\; 1 \leq i \leq N_Q
\end{equation}
Here, the $T_j$ are user-specified resolved basis functions, depending on the resolved solution $\mathbf{v}$. The $c_{i,j}$ are determined from the orthogonality constraints in \eqref{eq: orthogonality}, additionally requiring $c_{ii} = 1$. We choose $T_j = V_j$. This simple choice of basis functions has given good results in previous work on two-dimensional turbulence \cite{edeling_reducing_2020, hoekstra2024Reduced_data-driven}.

\subsection{Predictor-corrector setup}

The TO method is implemented in a predictor-corrector framework, where the low-fidelity solver first advances the system without the SGS term. The intermediate state, $\mathbf{v}^{n^*}$, is then corrected using the TO model:

\begin{align}
    \mathbf{v}^{n^*} &= S(\mathbf{v}^{n-1}), \nonumber\\
    \mathbf{v}^{n} &= \mathbf{v}^{n^*} + \int_{t^n-1}^{t^{n}} \mathbf{m}(\mathbf{v}^{n^*}) \D t, \nonumber\\
    &= \mathbf{v}^{n^*} + \mathbf{M}(\mathbf{v}^{n^*}),
\end{align}

where $S$ denotes the low fidelity solver, and $\mathbf{M}$ represents a model for the correction term. 

\paragraph{Tracking}
The predictor-corrector split allows the low-fidelity simulation to track a reference trajectory of QoIs, $\mathbf{Q}^\text{ref}(t^n)$, obtained from high-fidelity data. Given the predicted QoIs, $Q_i(\mathbf{v}^{n^*})$, we determine $\tau_i(t^n)$ from \eqref{eq: closed ODEs}:

\begin{align}
    \tau_i(t^n) &= \frac{\mathrm{d} Q_i^u}{\mathrm{d} t} \Big/ \int_\Omega \bar{P} \mathbf{V}_i \cdot R_i \mathbf{O}_i \D \boldsymbol{x}, \nonumber\\
    \frac{\mathrm{d} Q_i^u}{\mathrm{d} t} &= \frac{1}{\Delta t} \left(Q^\text{ref}_i(t^n) - Q_i(\mathbf{v}^{n^*})\right). \label{eq: dQ from ref}
\end{align}

This nudging approach ensures that the low-fidelity solver remains dynamically consistent with the reference trajectory while allowing us to extract statistical properties of the SGS term.

\paragraph{Predicting corrections}
An effective subgrid scale model should predict the SGS term without relying on reference trajectories. To this end, we train multivariate time-series models to predict the corrections based on historical data. These models learn the relationship between past QoI states and the unresolved-scale corrections.

\begin{figure}
    \centering

\usetikzlibrary{fit, positioning, backgrounds}

    \begin{tikzpicture}[
        boxnode/.style={draw, align=center, minimum width=2cm, minimum height=1cm},
        arrow/.style={->, thick},
        yellowarrow/.style={->, thick, Sepia},
        redarrow/.style={->, thick, red},
        labelstyle/.style={midway, font=\small, right}
        ]
        
        \node[boxnode] (qstar_n2) at (0,2) {$q^{n-2^*}$};
        \node[boxnode] (q_n2) at (0,0) {$q^{n-2}$};
        \node[boxnode] (qstar_n1) at (3,2) {$q^{n-1^*}$};
        \node[boxnode] (q_n1) at (3,0) {$q^{n-1}$};
        \node[boxnode] (qstar_n) at (6,2) {$q^{n^*}$};
        \node[boxnode] (q_n) at (9,0) {\textcolor{red}{$q^n$}};

        \draw[arrow] (qstar_n2) -- (q_n2) node[labelstyle] {$dQ^{n-2}$} ;
        \draw[arrow] (qstar_n1) -- node[labelstyle] {$dQ^{n-1}$} (q_n1);
        \draw[redarrow] (qstar_n) -- node[midway, right, red] {$dQ^n$} (q_n);
        
        \draw[yellowarrow] (q_n2) -- (qstar_n1);
        \draw[yellowarrow] (q_n1) -- (qstar_n) node[midway, right, Sepia]{LF solver};
        
        \node[draw, dashed, thick, inner sep=10pt, fit=(qstar_n2) (q_n1)] (modelhist) {};
        \node[above right] at (modelhist.south east) {Model hist};
        
        \begin{scope}[on background layer]
            \node[draw=blue, fill=blue!20, thick, inner sep=15pt, fit=(q_n2) (qstar_n)] (modelinput) {};
        \end{scope}
        \node[above right, blue] at (modelinput.north west) {LinReg input};
        

    \end{tikzpicture}

    \caption{Predictor-corrector subgrid scale term setup. During tracking, the corrections, $dQ$'s, are computed using Equation~\eqref{eq: dQ from ref}. We train models to predict either $dQ^n$ or $Q^n$ from historical inputs.}
    \label{fig: pred-cor sgs model setup}
\end{figure}
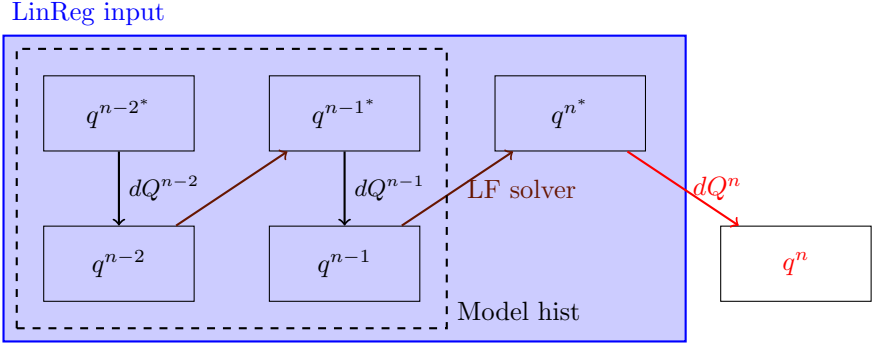

The overall setup is summarized in Figure~\ref{fig: pred-cor sgs model setup}, illustrating the correction process within the predictor-corrector framework. Here we introduce $dQ_i^n$ as the SGS correction to the $i$-th QoI at time $t^n$.

\subsection{Time-series models for SGS correction} \label{section: Time-Series Models for SGS Correction}
To model the unresolved dynamics, we adopt stochastic time-series models. Both the Mori-Zwanzig formalism \cite{franzke_stochastic_2016} and the ideal LES framework \cite{langford1999IDEALLES} emphasize that, in a coarse-grained simulation, a single resolved state may correspond to multiple high-fidelity realizations. This inherent uncertainty suggests that the SGS correction should be treated probabilistically.

Thus, rather than modeling a deterministic mapping from $\mathbf{v}^{n^*}$ to $dQ_i^n$, we construct a distribution of possible corrections. Advancing the simulation then involves sampling from this learned distribution, effectively introducing a stochastic SGS term.

We explore two modeling approaches for predicting the SGS correction: a simple data-driven noise model and a linear regression model:
\begin{itemize}
    \item \begin{bf}Data-driven noise model (DDM)\end{bf} 
    
    In previous work \cite{hoekstra2024Reduced_data-driven}, we used a multivariate Gaussian model trained on the $dQ$ corrections obtained from tracking. At each time step, the SGS correction was drawn from this learned distribution. While simple, this model does not account for temporal correlations or even a dependence on the current state of the system.
    \item \begin{bf}Linear regression with stochastic residuals (LRS)\end{bf}

    To improve predictive accuracy, we extend the data-driven noise model with a linear regression term. This model predicts the corrected QoIs using a history of previous states:   
    \begin{align}
        \mathbf{LRS}(\underline{\mathbf{q}}_h) = \underline{\mathbf{q}}_h C + \boldsymbol{\eta}, \quad \boldsymbol{\eta} \sim \mathcal{N}(\boldsymbol{\mu}, \Sigma), \nonumber\\
        \underline{\mathbf{q}}_h = [\mathbf{q}^{n-1}, \dots, \mathbf{q}^{n-h}, \mathbf{q}^{n^*}, \mathbf{q}^{n-1^*}, \dots, \mathbf{q}^{n-h^*}, 1], \label{eq: Linear Regression with Stochastic Residuals}
    \end{align}
    where $\underline{\mathbf{q}}_h$ is a vector with all model inputs for history length $h$, $C \in \mathbb{R}^{ len(\underline{\mathbf{q}}_h) \times N_Q}$ is a learnable regression matrix, and $\mathcal{N}$ is a $N_Q$-dimensional multivariate Gaussian distribution. The ``1" in the model inputs allows us to learn a bias term. We fit the model to predict the corrected QoIs $\mathbf{q}^{n}$. 
    We solve a regularized least squares problem to find $C$. Where we regularize using the Frobenius-norm of $C$ to promote long-term stability of the coupled LES-SGS system, since the Frobenius-norm penalizes large singular values:
    \begin{equation}
        C = \argmin_{X \in \mathbb{R}^{ len(\underline{\mathbf{q}}_h) \times N_Q}} \frac{1}{2} \lVert \mathbf{Q}_h X - \mathbf{Q} \rVert_F^2 + \lambda \lVert X \rVert_F,
    \end{equation}
    where the $n$-th rows of $\mathbf{Q}_h$ and $\mathbf{Q}$ are $\underline{\mathbf{q}}_h^n$ and $\mathbf{q}^{n}$ from the training data points, and $\lambda$ is the regularization parameter. Before fitting the model, we standardize the inputs and outputs by dividing by the standard deviation of the reference trajectories.
\end{itemize}

\subsection{Final integration into LES}
By replacing explicit tracking with self-generated predictions for $dQ_i$, the final model achieves a fully independent SGS closure. The model has between 100 and 1000 tunable parameters, 
depending on the number of QoIs and the history length in the linear regression. This makes it computationally cheaper than traditional deep learning based SGS models, which can easily contain many millions of tunable constants. Moreover, the time series model predicts corrections to the QoIs which are more interpretable.

\section{Turbulence in a box} \label{section 4: HIT}

To evaluate the performance of our reduced SGS model and find suitable hyper parameter settings we first conduct long-term simulations of three-dimensional homogeneous isotropic turbulence (HIT) in a periodic box.

\subsection{Forcing}
Sustaining isotropic turbulence requires external forcing. The choice of forcing scheme influences large-scale flow structures and the resulting turbulence spectrum. A simple choice, forcing only one wavenumber in one direction, results in Kolmogorov flow \cite{shebalin_kolmogorov_1997}. However, this introduces anisotropy in the large scales. 

To avoid this, we adopt a stochastic forcing scheme based on Ornstein-Uhlenbeck (OU) processes \cite{eswaran_examination_1988}, following the efficient implementation by \cite{chouippe_forcing_2015}. The forcing is applied in Fourier space,
where only modes with wavenumber magnitude $\lVert \boldsymbol{k} \rVert \leq C_F$ are excited. Each forced Fourier coefficient $\hat{\boldsymbol{f}}_{\boldsymbol{k}}$ evolves independently as:

\begin{equation}
    \hat{\boldsymbol{f}}_{\boldsymbol{k}}(t+\Delta t_f) = \hat{\boldsymbol{f}}_{\boldsymbol{k}}( t)\left(1 - \frac{\Delta t_f}{T_L}\right) + W(t)\left(2 \sigma^2 \frac{\Delta t_f}{T_L}\right)^{1 / 2},
\end{equation}

where $\Delta t_f$ is the forcing time step, $T_L$ is the characteristic timescale, $W(t)$ denotes white noise, and $\sigma^2$ controls the variance. Instead of directly prescribing $\sigma$, we specify the energy injection rate as $e^* = \sigma^2 T_L$, see Table \ref{tab:DNS parameters}. The forcing field in physical space is obtained by an inverse Fourier transform and subsequently projected to ensure incompressibility.

\subsection{High-fidelity simulation}
To generate reference data, we performed a high-fidelity simulation at $N = 512^3$, with a time step of $\Delta t = 2.5 \cdot 10^{-4}$. The forcing term was updated every 10-th time step, to enable consistent forcing on low-fidelity simulations. Table \ref{tab:DNS parameters} summarizes the key parameters.

\begin{table}[h]
    \centering
    \begin{tabular}{c|c|c|c|c|c|c}
        $N$ & $\Delta t$ & $1/\nu$           & $T_L$ & $\Delta t_f$  & $e^*$ & $k_f$\\ \hline
        $512^3$ & $2.5 \cdot 10^{-4}$ & $2000$ & $0.01$ & $2.5 \cdot 10^{-3}$ & $0.1$ & $\sqrt{2}$
    \end{tabular}
    \caption{High-fidelity parameters HIT (homogeneous isotropic turbulence).}
    \label{tab:DNS parameters}
\end{table}

\begin{figure}[]
    \centering
    \begin{minipage}[b]{0.45\textwidth}
    \includegraphics[width=\linewidth]
      {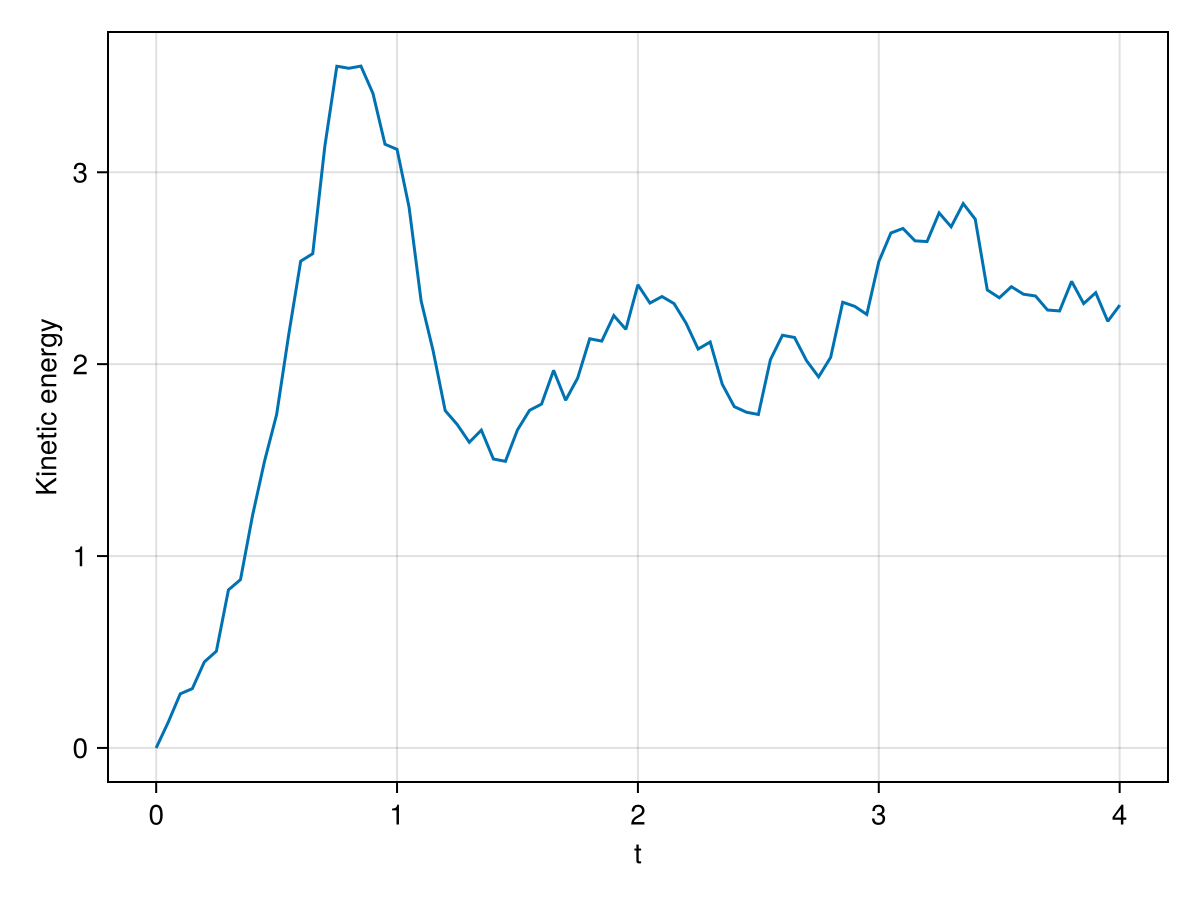}
      \caption{\centering Kinetic energy during spin-up of high-fidelity solver.}
      \label{fig: kenitic energy during spinup}
    \end{minipage}
    \hfill
    \begin{minipage}[b]{0.5\textwidth}
      \includegraphics[width=\linewidth]{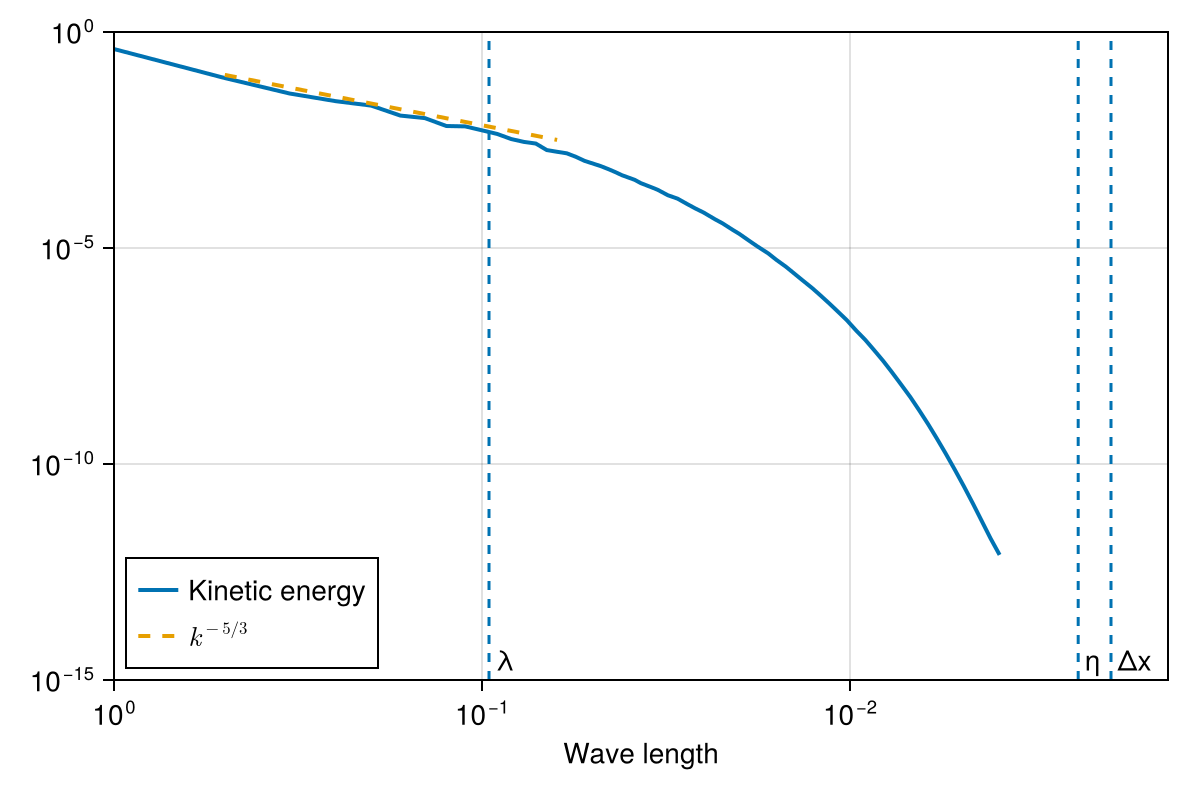}
      \caption{\centering Energy spectrum at $t=4$ in high-fidelity solver.}
      \label{fig: spectrum initial field}
    \end{minipage}
  \end{figure}

The velocity field was initialized at rest. Turbulence developed over a spin-up period of four time units. Figure \ref{fig: kenitic energy during spinup} shows the evolution of kinetic energy, illustrating how energy accumulates before reaching a statistical steady state. The flow statistics at $t=4$ are listed in Table \ref{tab: turbulence_statistics_initial_field}, and the corresponding energy spectrum is shown in Figure \ref{fig: spectrum initial field}. In the figure, we included the Kolmogorov length scale, the Taylor length scale and the grid size. Note that the Kolmogorov length scale is limited by the grid size, simulations at this resolution should therefore not be seen as a full DNS. However, this resolution does lead to a nice inertial range and was used in this paper to run ground-truth simulations over 100 time units.

\begin{table}[h]
    \centering
    \begin{tabular}{c|c|c|c|c|c|c|c}
        $u_\text{avg}$ & $\epsilon$ & $\eta$ & $\lambda$ & $L$ & $Re_\text{int}$ & $Re_\lambda$ & $t_\text{int}$  \\ \hline
        3.72 & 3.78 & 0.0024 & 0.096 & 0.15 & 1100 & 411 & 0.04
    \end{tabular}
    \caption{Turbulence statistics at the end of spin-up: Mean velocity, dissipation rate, Kolmogorov length scale, Taylor length scale, integral length scale, Reynolds number, Taylor-scale Reynolds number, large-eddy turnover time.}
    \label{tab: turbulence_statistics_initial_field}
\end{table}

\subsection{Low-fidelity simulation}
\begin{figure}
    \centering
    \includegraphics[width=0.7\linewidth]{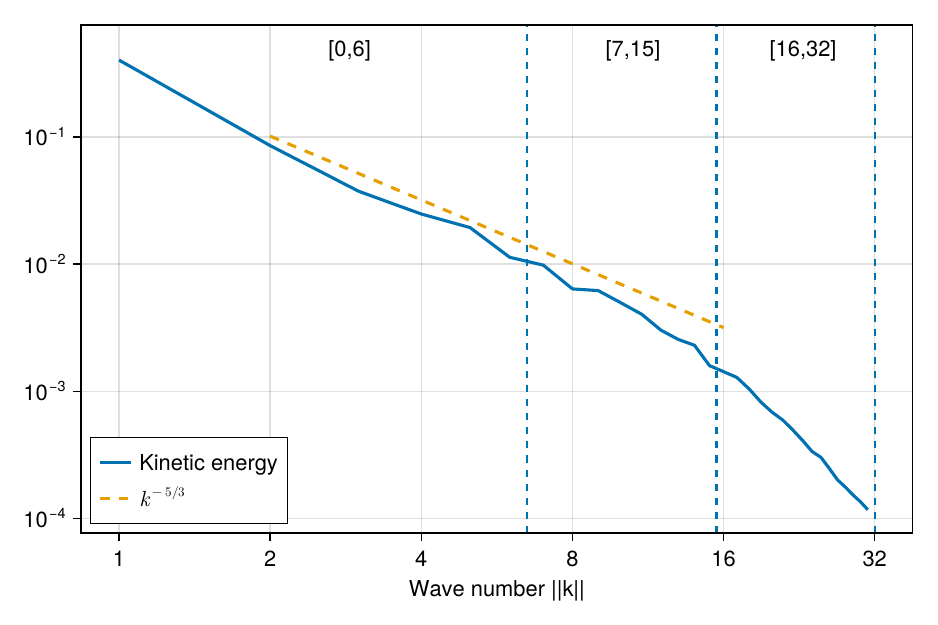}
    \caption{Energy spectrum coarse-grained initial field.}
    \label{fig: coarse-grained initial spectrum}
\end{figure}

For the low-fidelity (LF) simulations, we used a solver with resolution $N = 64^3$ and a time step ten times larger than that of the HF simulation, $\Delta t_\text{LF} = 2.5 \cdot 10^{-3}$. Figure~\ref{fig: coarse-grained initial spectrum} displays the energy spectrum of the coarse-grained initial velocity field, taken from the HF simulation at $t = 4$.

\begin{figure}[]
    \centering
      \includegraphics[width=0.7\linewidth]{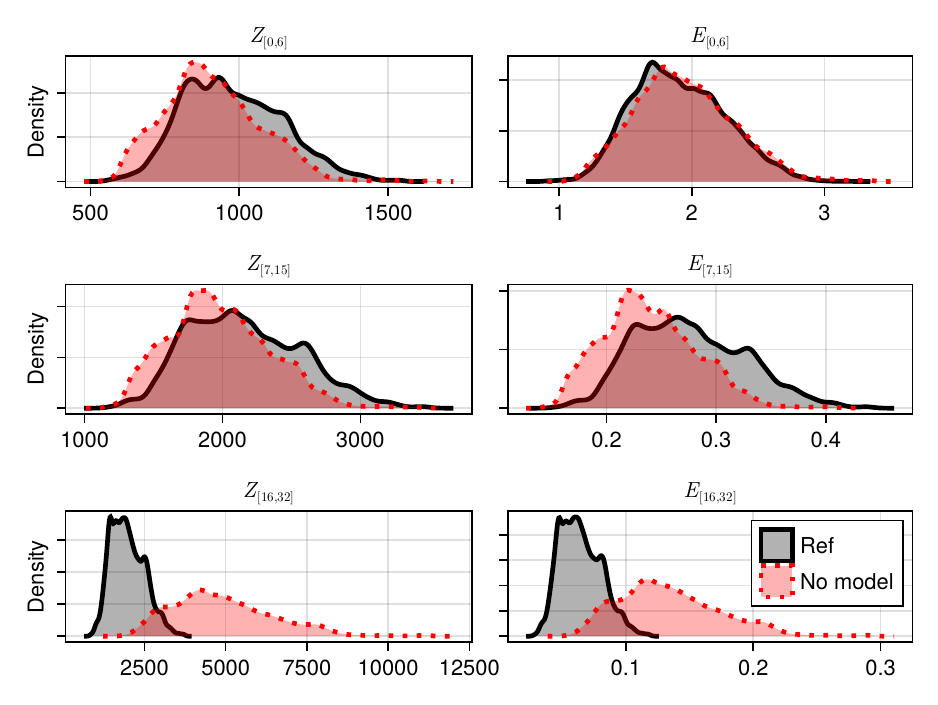}
    \caption{Long-term QoI distributions LF solver without SGS term.}
    \label{fig: long-term distributions no model}
\end{figure}

To assess the impact of different SGS models, we analyze six QoIs: scale-aware energy and enstrophy in three wavenumber bands: $[0,6]$, $[7,15]$, and $[16,32]$. Figure~\ref{fig: long-term distributions no model} shows the long-term distributions of these QoIs over 100 time units for a baseline LF simulation without an SGS model, compared against the HF reference data. The largest discrepancies are in the high wavenumber energy $E_{[16,32]}$ and enstrophy $Z_{[16,32]}$, where the long-term distributions shifted to the right, indicating accumulation of energy near the cutoff scale of the LF simulation. 

To quantify such discrepancies throughout the paper, we use the Kolmogorov–Smirnov (KS) distance, which compares two distributions by measuring the maximum difference between their cumulative density functions $F(x)$ and $G(x)$:
\begin{equation}
    KS\Big(F(x),G(x)\Big) = \max_x \Big\lvert F(x)-G(x) \Big\rvert.
    \label{eq:KS}
\end{equation}
The KS-distance is especially well-suited for this context because it does not require integration over the distributions’ support, making it robust when comparing QoIs with different magnitudes or units. This allows us to summarize discrepancies across all QoI distributions in one number; the summed KS-distance. Table \ref{tab: KS-distances LF-solver} reports the KS-distances between the long-term distributions for the QoIs in the low-fidelity and high-fidelity solver.
\begin{table}[]
    \centering
    \begin{tabular}{c|c|c|c|c|c|c|c}
         & $Z_{[0,6]}$ & $E_{[0,6]}$ & $Z_{[7,15]}$ & $E_{[7,15]}$ & $Z_{[16,32]}$ & $E_{[16,32]}$ & sum \\ \hline
         No model &  0.16 & 0.047 & 0.17 & 0.29 & 0.87 & 0.80 & 2.34 \\
         Smag 0.071 &  0.13 & 0.078 & 0.22 & 0.19 & 0.067 & 0.021 & 0.705
    \end{tabular}
    \caption{KS-distance between long-term distributions in HF simulation and LF simulation with or without Smagorisky model.}
    \label{tab: KS-distances LF-solver}
\end{table}

\begin{figure}
    \centering
    
\end{figure}

\subsection{Smagorinsky model}
The Smagorinsky model \cite{smagorinsky_general_1963} addresses the energy accumulation at high wavenumbers of the LF simulation by introducing an eddy-viscosity term based on the local strain rate:

\begin{align}
    \mathbf{m}(\mathbf{v}, C_s) &= \nabla \cdot (2 \nu_t S_{ij}), \\
    \nu_t &= C_s^2 \Delta^2 \sqrt{2 S_{ij}S_{ij}}.
\end{align}

\begin{figure}
    \centering
    \includegraphics[width=0.7\linewidth]{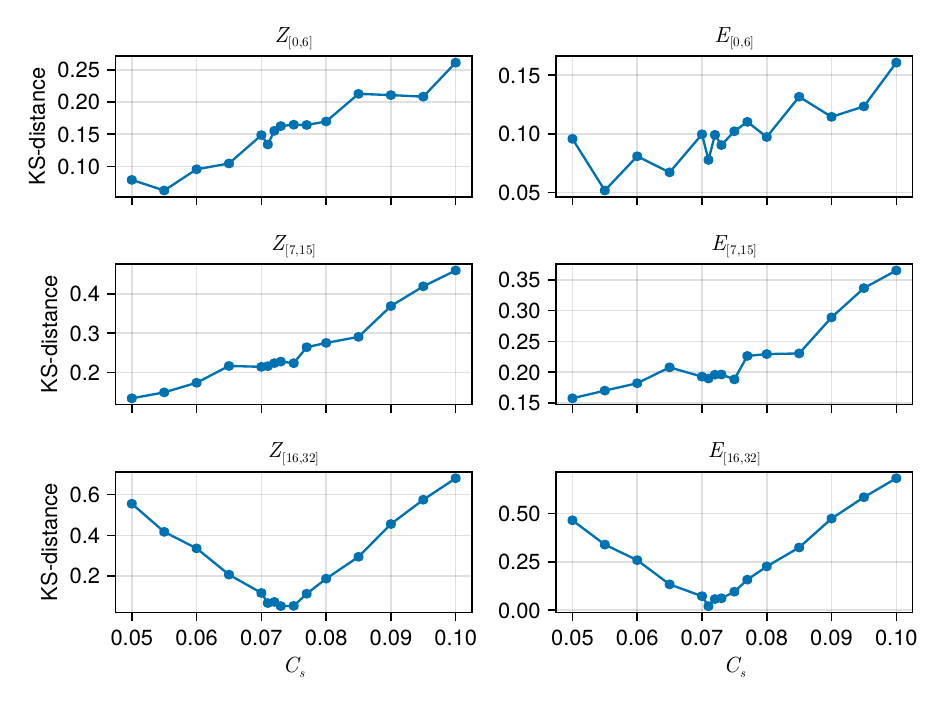}
      \caption{Distance between 100 time unit QoI distributions for different values of $C_s$.}
      \label{fig: tune cs}
\end{figure}

\begin{figure}
    \centering
      \includegraphics[width=0.7\linewidth]{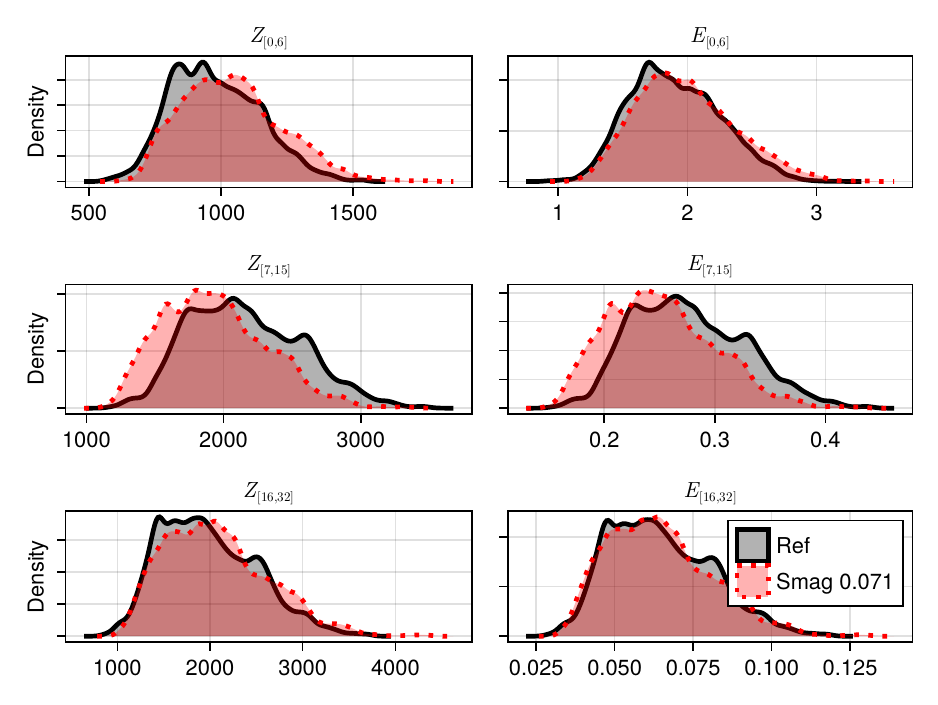}
      \caption{Long-term QoI distributions with Smagorinsky model.}
      \label{fig: long-term smag}
\end{figure}

Here, $S_{ij} = 1/2(\partial v_i/ \partial x_j + \partial v_j/ \partial x_i)$ is the strain rate tensor, $\Delta$ is the filter width (set to the LF grid size), and $C_s$ is the Smagorinsky constant. We tuned $C_s$ for best agreement with the reference distributions, as shown in Figure \ref{fig: tune cs}. The best agreement, in terms of the summed KS-distance across all QoIs, was obtained with $C_s = 0.071$, yielding a summed KS-distance of 0.705. The corresponding long-term distributions are plotted in Figure~\ref{fig: long-term smag}.

\begin{figure}[]
    \centering
      \includegraphics[width=0.7\linewidth]{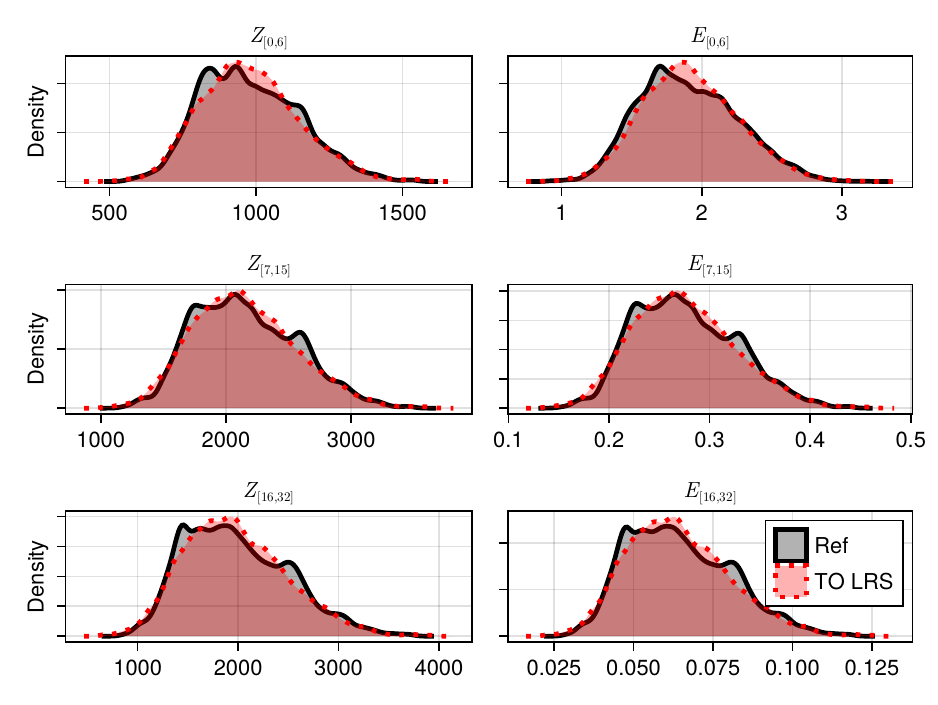}
    \caption{Long-term QoI distributions of 5-member ensemble with TO SGS term, using LRS with history length 5.}
    \label{fig: long-term distributions TO}
\end{figure}

\subsection{TO model}
We trained the TO method using time series of the six QoIs from the HF simulation. Specifically, we used the first 10 time units of data to train the models and evaluated their performance over 100 time units. Training data for $dQ$ was obtained by running a 10 time unit LF simulation that tracked the reference trajectories via equation~\eqref{eq: dQ from ref}. The effect of reducing the amount of training data is explored in Section~\ref{section: TO on small data}.

Subsequently, we fitted linear regression models with stochastic residuals (LRS) using different history lengths. We also trained a data-driven noise model (DDN), which only fits a multivariate Gaussian distribution to the $dQ$ data. These models were evaluated as subgrid scale term in LF simulations over 100 time units. Since these models are stochastic, we ran five replica simulations of each model with a different random seed.

\begin{figure}
    \centering
    \includegraphics[width=0.8\linewidth]{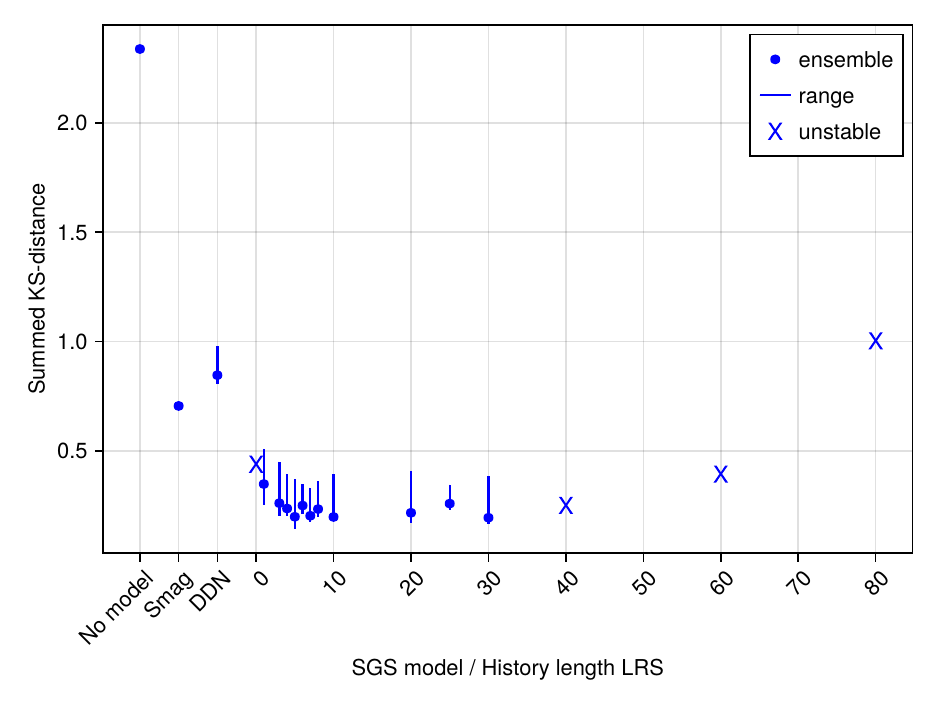}
    \caption{Predictive quality of SGS models, extrapolating from a 10 time unit training domain to 100 time units. DDN refers to the data-driven noise model, LRS refers to the linear regression with stochastic residuals, and smag is the Smagorinsky model with $C_s = 0.071$. The plotted ranges are based on 5 replica simulations. The ensemble KS-distance results from aggregating the QoI data of all ensemble members. ``X'' markers indicate at least one unstable replica, their location is based on the ensemble KS-distance of the stable part of the trajectories.}
    \label{fig:Hit large data online}
\end{figure}

Figure \ref{fig:Hit large data online} reports the KS-distances between the resulting long-term QoI distributions and the HF reference. We show the range between the KS-distance of the worst and best replica as well as the ensemble KS-distance, which results from aggregating the QoI data of all ensemble members in one EDF (empirical density function). Models leading to unstable trajectories in at least one replica are marked with an ``X". The DDN model performs slightly worse than the optimized Smagorinsky model. In contrast, the LRS models exhibit good performance, achieving near-optimal accuracy across a wide range of history lengths. This indicates that the method is not highly sensitive to this hyperparameter. Instabilities are observed only for models with zero history (i.e., relying only on $\mathbf{q}^{n^*}$) or very long history lengths.

Figure \ref{fig: long-term distributions TO} contains the long-term distributions of one of the best performing models: LRS with history length 5. Here we plotted the histogram which combines the QoI trajectories of all five ensemble members.

All LRS models in this section were trained without regularization. \ref{app: regularization} shows that adding L2 regularization improves stability for long history lengths, but at the cost of degraded long-term distribution accuracy.

\subsection{Beyond QoI distributions}
Thus far, we have shown that the TO method, combined with a linear regression model using a history length of five, accurately reproduces the long-term statistics of the QoIs it was trained on. In this section, we extend the analysis by examining segments of the resulting short-term QoI trajectories and evaluating additional turbulence characteristics; the energy spectrum and coherent flow structures.

\begin{figure}[]
    \centering
      \includegraphics[width=0.7\linewidth]{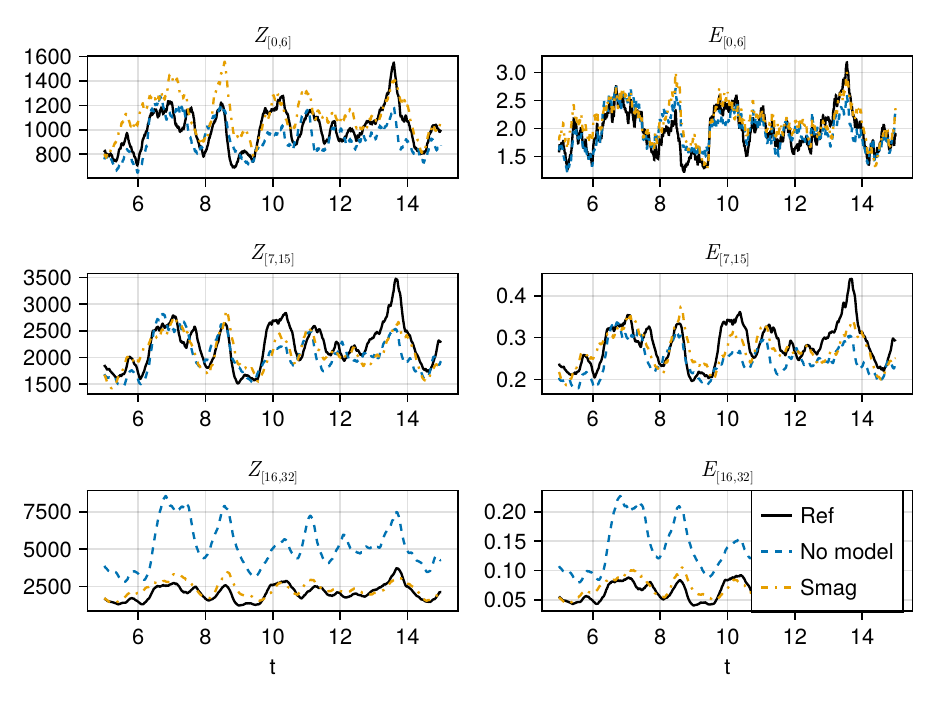}
    \caption{QoI trajectories without SGS model and with optimized Smagorinsky model.}
    \label{fig: qoi trajectories nomodel smag}
\end{figure}

\begin{figure}
    \centering
      \includegraphics[width=0.7\linewidth]{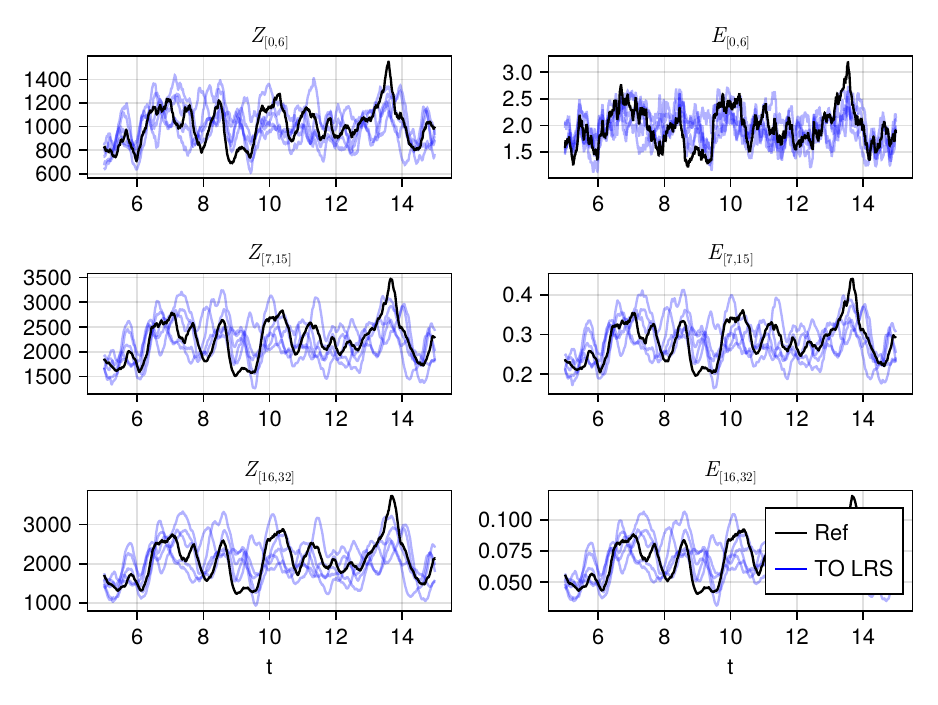}
      \caption{QoI trajectories with TO LRS model with history length 5.}
      \label{fig: qoi trajectories TO}
\end{figure}

Figures \ref{fig: qoi trajectories nomodel smag} and \ref{fig: qoi trajectories TO} show short segments of the QoI trajectories for different SGS models. In Figure \ref{fig: qoi trajectories nomodel smag}, the no model case clearly deviates from the reference across most QoIs. The Smagorinsky model reduces this discrepancy, although some deviation in amplitude remains. Figure \ref{fig: qoi trajectories TO} displays the TO model with history length 5, plotted as five ensemble members. While the TO model generally follows the reference trajectory more closely and preserves variability in most bands, occasional underestimation of amplitudes shows remaining shortcomings in its predictive accuracy.

\begin{figure}
    \centering
    \includegraphics[width=0.7\linewidth]{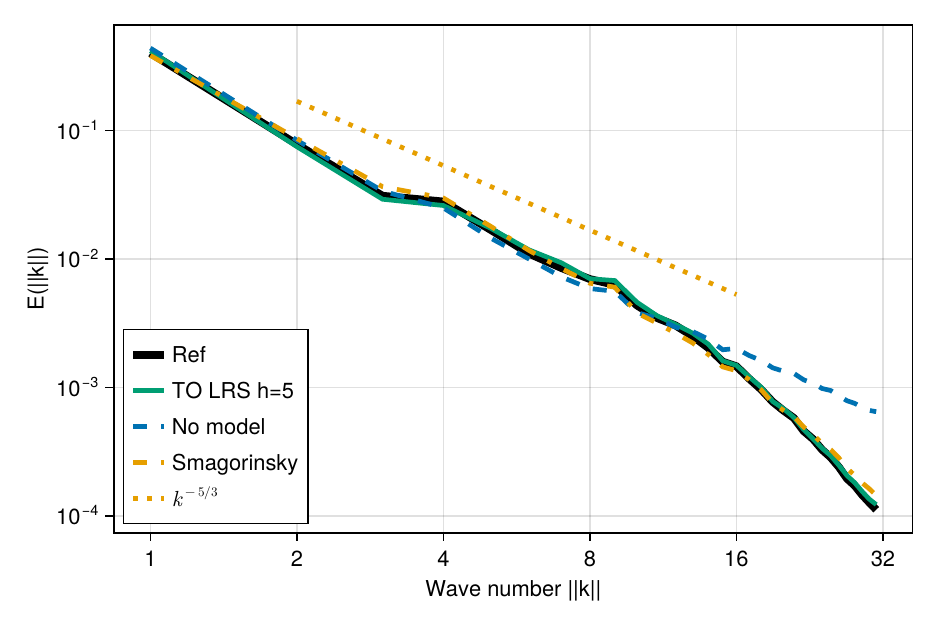}
    \caption{Energy spectrum for various SGS models.}
    \label{fig:energy spectra online}
\end{figure}

Figure \ref{fig:energy spectra online} presents the time-averaged energy spectrum, computed over 10 snapshots taken between $t=75$ and $t=100$ in the long-term simulations. For the TO model, we show results from the first ensemble member. The LF simulation without an SGS model shows clear energy accumulation at high wavenumbers. The TO model achieves a slightly improved match to the HF reference spectrum compared to the Smagorinsky model.

\begin{figure}
    \begin{subfigure}[b]{0.3\textwidth}
        \centering
        \includegraphics[width = \linewidth]{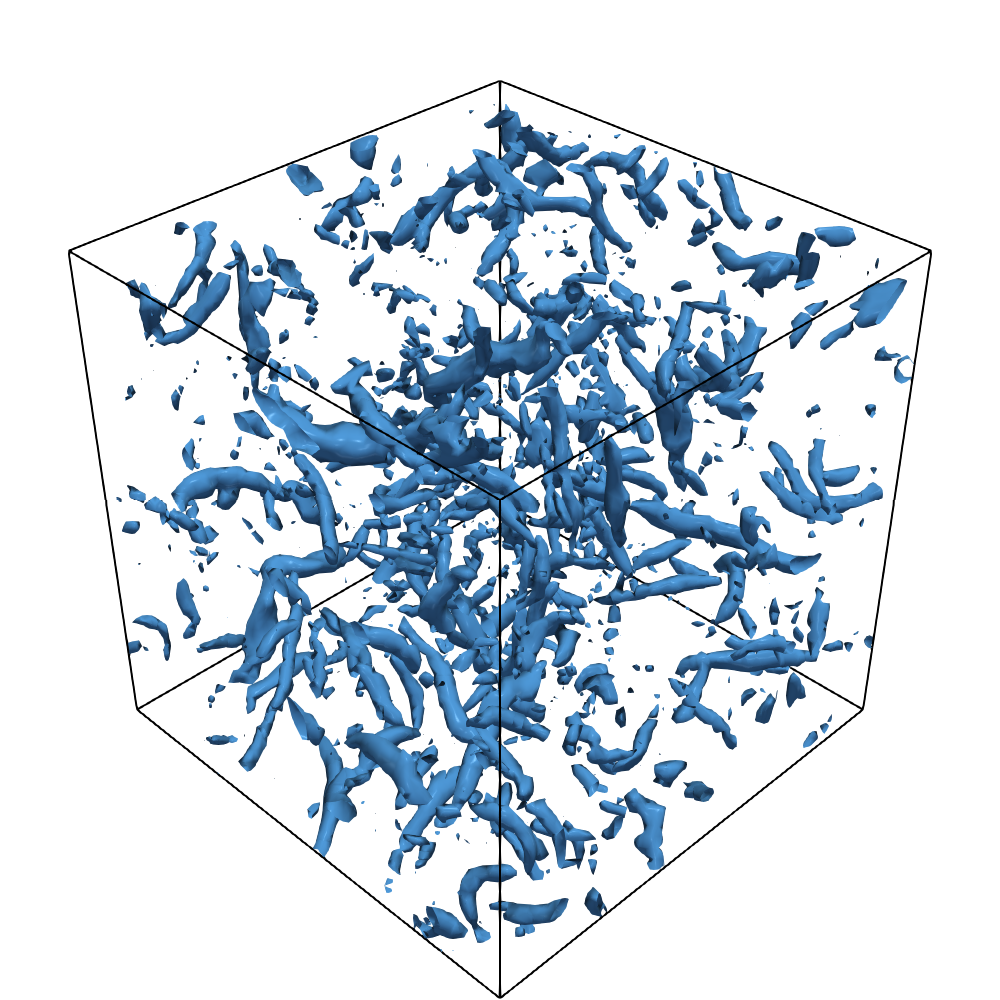}
        \caption{Filtered HF}
    \end{subfigure}
    \begin{subfigure}[b]{0.3\textwidth}
        \centering
        \includegraphics[width = \linewidth]{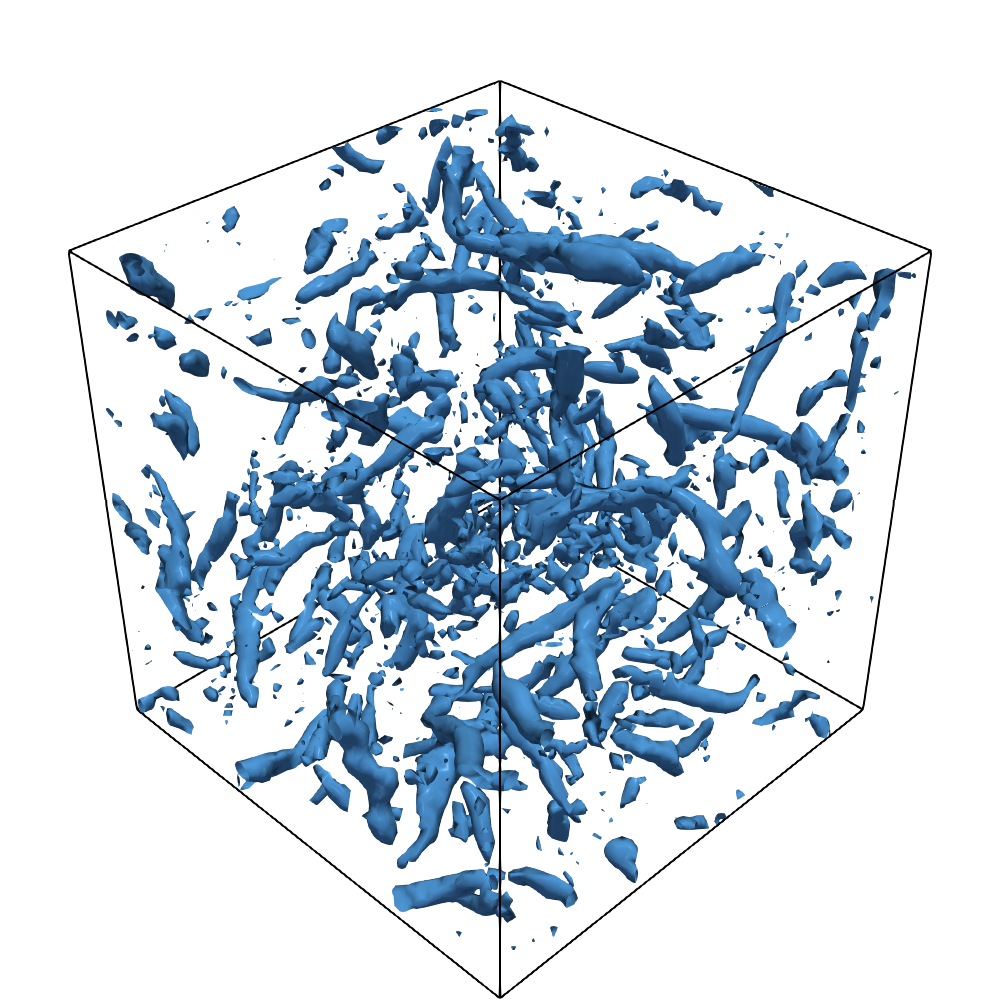}
        \caption{TO LRS h=5}
    \end{subfigure}
    \begin{subfigure}[b]{0.3\textwidth}
        \centering
        \includegraphics[width = \linewidth]{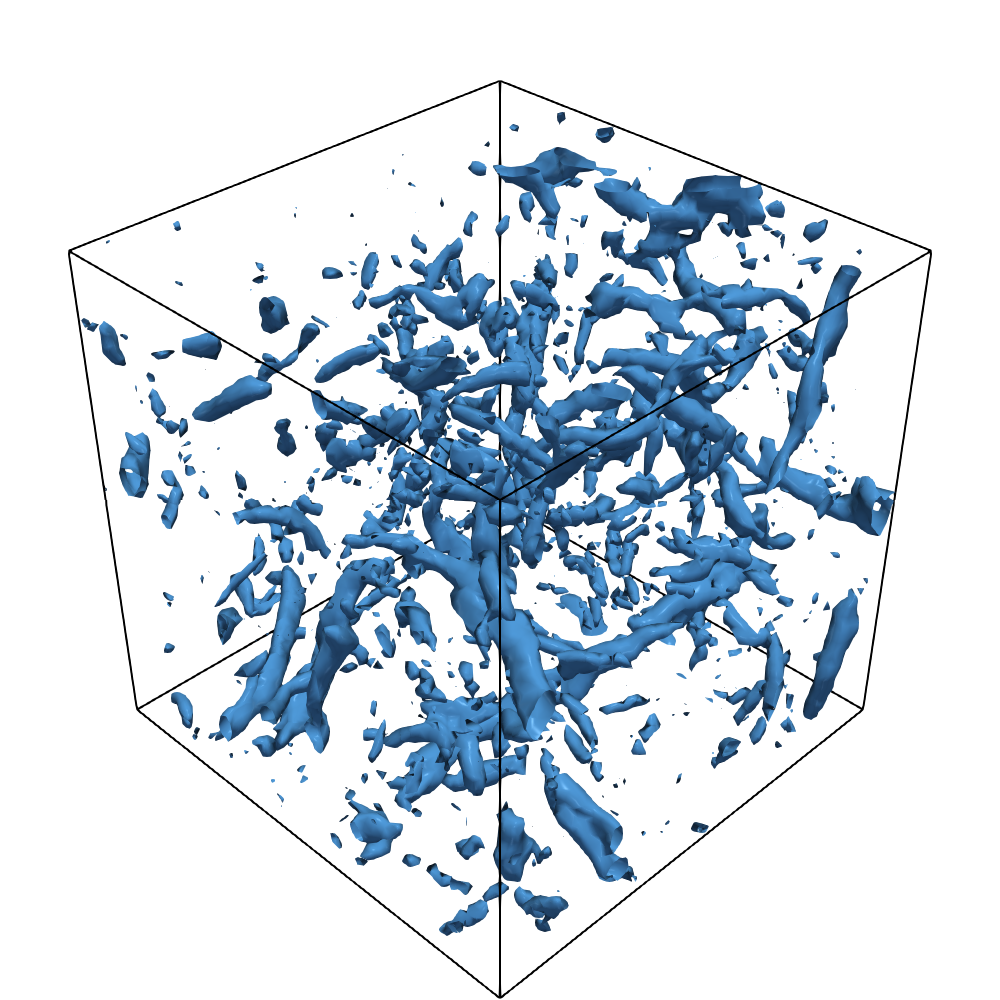}
        \caption{Smag 0.071}
    \end{subfigure}
    \begin{subfigure}[b]{0.3\textwidth}
        \centering
        \includegraphics[width = \linewidth]{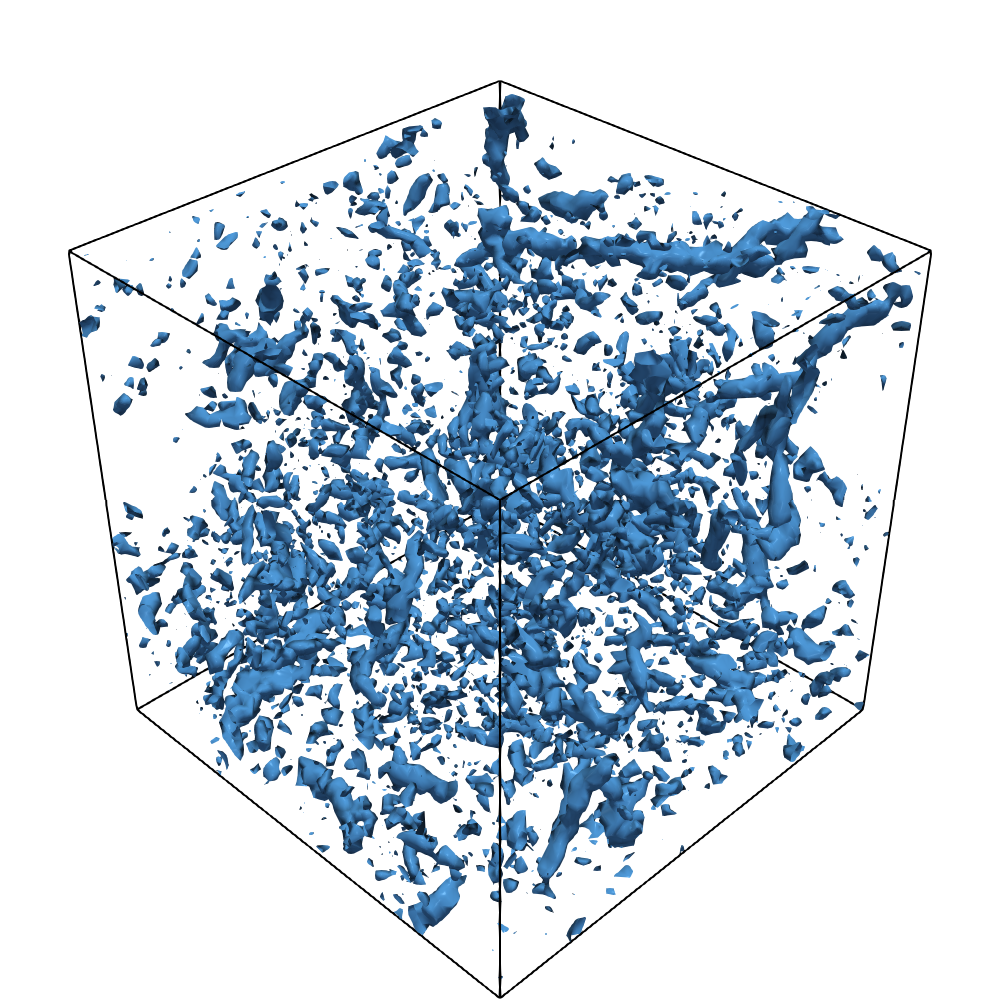}
        \caption{No model}
    \end{subfigure}
    \caption{Turbulent vortices at $T=100$ in long-term simulations, visualized via isocountours $Q = 2000$. Here $Q = \frac{1}{2}(\lVert \Omega \rVert^2 - \lVert S \rVert^2)$, and can be seen as the local balance between vorticity magnitude and shear strain rate.}
    \label{fig:Turbulent structures final fields}
\end{figure}

Figure \ref{fig:Turbulent structures final fields} visualizes the turbulent structures in the final fields of the long-term simulations using isocontours of the second invariant of the velocity gradient tensor, a standard method for identifying coherent vortices \cite{lesieur_large-eddy_2005}. Both the TO and Smagorinsky models preserve these structures, while the simulation without an SGS model exhibits smaller, more numerous vortices.


\subsection{Performance of TO LRS on limited training data} \label{section: TO on small data}
To evaluate the data efficiency of the TO LRS method, we examined how model performance varies with the amount of available training data. Specifically, we trained models using history lengths of 5 and 10, across a range of training data sizes. We also explored the effect of different regularization strengths on performance. 

\begin{figure}[h]
    \centering
    \begin{subfigure}[b]{0.40\textwidth}
        \centering
        \includegraphics[width = \linewidth]{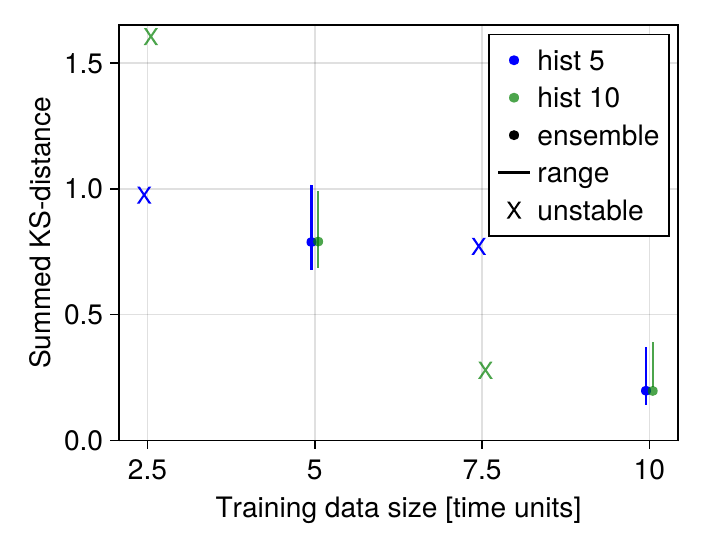}
        \caption{$\lambda = 0$}
    \end{subfigure}
    \begin{subfigure}[b]{0.40\textwidth}
        \centering
        \includegraphics[width = \linewidth]{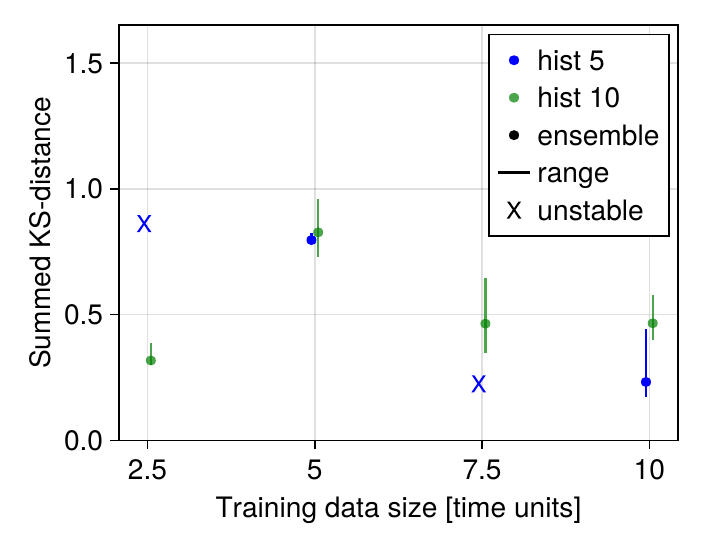}
        \caption{$\lambda = 0.01$}
    \end{subfigure}
    
    \begin{subfigure}[b]{0.40\textwidth}
        \centering
        \includegraphics[width = \linewidth]{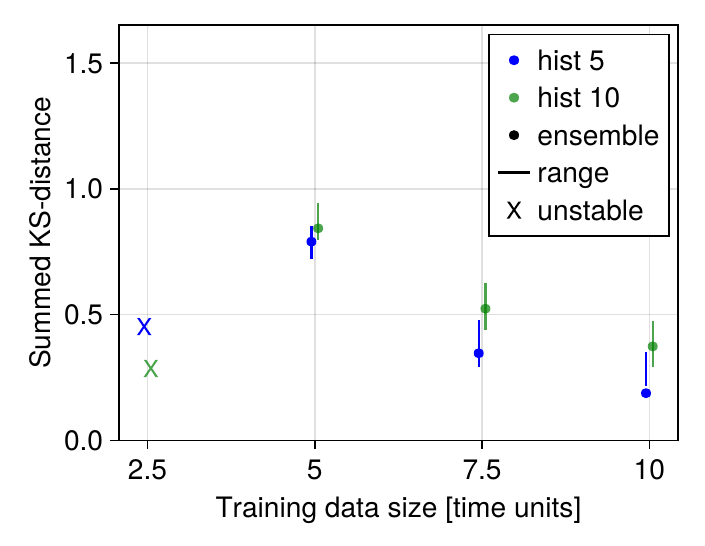}
        \caption{$\lambda = 0.1$}
    \end{subfigure}
    \caption{Predictive quality of the TO LRS model with less training data, for various regularization strengths. Ranges are based on 5 replica simulations. ``X'' markers indicate at least one unstable replica, their location is based on the ensemble KS-distance of the stable part of the trajectories.}
    \label{fig:small training data}
\end{figure}

The results of these experiments are presented in Figure \ref{fig:small training data}. We observe a decline in performance as the amount of training data decreases. Without regularization, models trained on 7.5 time units of data already exhibit instability in simulations. Introducing regularization mitigates this issue and results in a smoother dependence of performance on training data size.

A methodological note: we consistently excluded the initial unit of simulation time from the training set to avoid bias due to transient dynamics at the start of the low-fidelity simulations. As such, a model reported as trained on 2.5 time units effectively used data from $1 \leq t \leq 2.5$.

\section{Channel flow} \label{section 5: Channel}
To demonstrate the versatility of the TO LRS approach, we apply it to a turbulent channel flow. Channel flow is a classical benchmark in turbulence research due to its relatively simple geometry and the presence of well-known turbulent structures. However, it presents a significant modeling challenge due to the different turbulence layers. Recall that the TO LRS model relies solely on the history of QoIs and is largely agnostic to geometry. This makes channel flow a stringent test of its generalizability.

\subsection{Setup}
We ran simulations following the general setup described in \cite{vreman_comparison_2014}, who performed direct numerical simulations of turbulent channel flow at a friction Reynolds number of $Re_\tau = 180$. They compared these simulations to existing data bases, among which \cite{moser_direct_1999}. We will use publicly available data from \cite{vreman_comparison_2014} to verify our SGS models for this test case.

The channel has dimensions $L_x \times L_y \times L_z = 4\pi \times 2 \times 4/3\pi$, where $x$, $y$, and $z$ denote the streamwise, wall-normal, and spanwise directions, respectively. The domain is periodic in the streamwise and spanwise directions, with no-slip boundary conditions applied at the walls ($y = 0$ and $y=2$). A constant mean pressure gradient in the streamwise direction is imposed to drive the flow. We express $y$ in wall units $y^+ = 180y$.

\begin{figure}
\centering
    \begin{subfigure}[b]{0.7\textwidth}
        \centering
        \caption{High resolution.}
        \includegraphics[width = \linewidth]{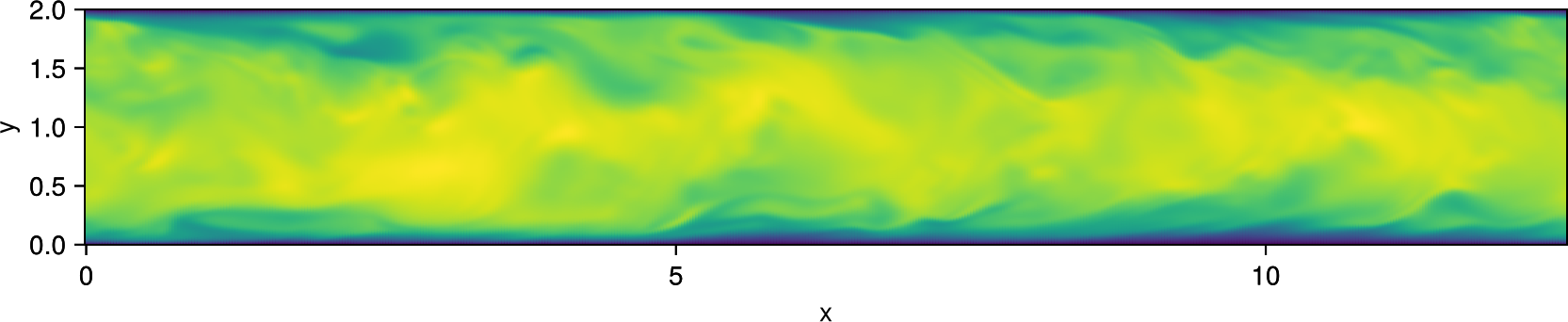}
        \label{fig: channel initial fine}
    \end{subfigure}
    \begin{subfigure}[b]{0.7\textwidth}
        \centering
        \caption{Low resolution (coarse-grained).}
        \includegraphics[width = \linewidth]{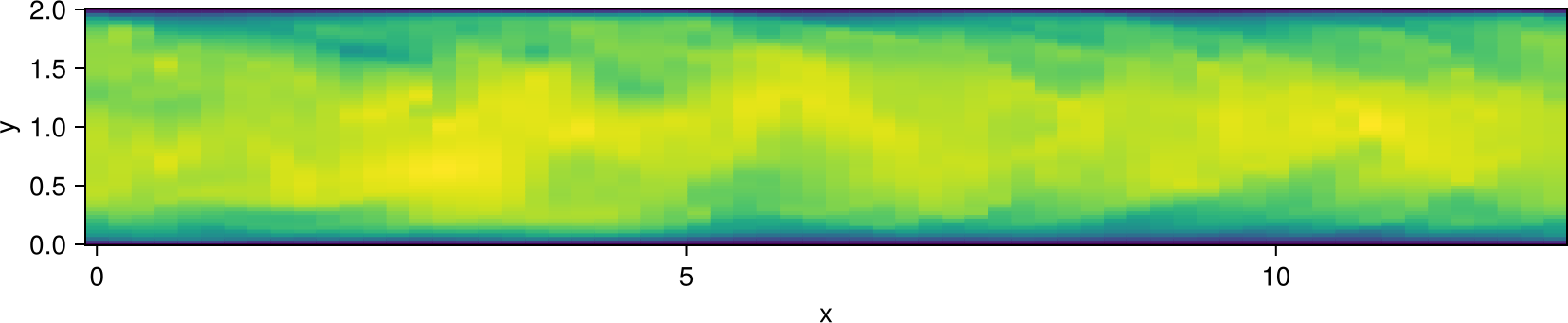}
        \label{fig: channel initial coarse}
    \end{subfigure}
    \caption{Initial turbulent field for channel flow experiments, cross section of x-velocity $\mathbf{u}_1$ at $z=0$}
    \label{fig: channel initial}
\end{figure}

We adapted the initial conditions from \cite{moin_numerical_1980} and used these in a 15 time unit spin-up simulation to create a turbulent initial field for our experiments, see figure \ref{fig: channel initial fine}.

\subsection{High-fidelity simulation}

To generate training data (QoI trajectories), we ran our own high-fidelity simulation, using a $512 \times 512 \times 256$ equidistance grid.
This grid choice yields a resolution comparable to the smallest cells used in the coarsest DNS of \cite{vreman_comparison_2014}, who employed a $256 \times 128 \times 128$ grid with a tangent hyperbolic stretching in the wall-normal direction.
The setup is summarized in table \ref{tab:channel setup}. We choose to use an equidistance grid to keep the coarse-graining operation and the expressions for the scale-aware QoIs simple. 

\begin{table}[H]
    \centering
    \begin{tabular}{c|c|c|c|c|c|c}
       $L_x \times L_y \times L_z$ & $N_x \times N_y \times N_z$ & $\Delta t$ & $Re_\tau$ & $\nu$ & $y^+$ & $\mathbf{f}$ \\ \hline
       $4\pi \times 2 \times 4/3\pi$& $512 \times 512 \times 256$ & $5 \cdot 10^{-4}$ & 180 & 1/180 & $180y$ & $(1, 0, 0)^T$
    \end{tabular}
    \caption{DNS parameters channel flow.}
    \label{tab:channel setup}
\end{table}

\subsection{Low-fidelity simulation}
The low-fidelity simulations were performed on a $64 \times 64 \times 32$ grid with a timestep of $\Delta t_{LF} = 5 \cdot 10^{-3}$. We used face-averaging to coarsen the HF initial velocity field to the LF initial field. This coarse-grained initial field is plotted in Figure \ref{fig: channel initial coarse}.

\subsubsection{Scale-aware QoIs}
As in the previous test case, we define six scale-aware QoIs based on energy and enstrophy content in three Fourier bands: $[0, 3]$, $[4, 10]$, and $[11, 17]$. To ensure periodicity in all directions (a requirement for Fourier analysis), we symmetrically extend the solution field in the wall-normal direction. Specifically, a vector field $\mathbf{f}$ on the domain $4\pi \times 2 \times 4/3\pi$ is extended to a periodic vector field $\tilde{\mathbf{f}}$ on $4\pi \times 4 \times 4/3\pi$ by mirroring in the wall at $y=2$:
\begin{equation}
    \tilde{\mathbf{f}}(x,y,z) = \begin{cases} \mathbf{f}(x,y,z), &\text{ if } y\leq 2, \\ 
    -\mathbf{f}(x,4-y,z), &\text{ if } y > 2.
    \end{cases}
\end{equation}
Due to the domain size, the resulting Fourier wavenumbers are fractional. For instance, the longest non-constant wave in the streamwise direction corresponds to a wavenumber of $1/(4\pi)$. To illustrate the spatial features captured by the different Fourier bands, we visualized the velocity and vorticity fields which result from filtering the initial field with the different scale-aware filters in \ref{app: Scale-aware filters applied to initial field channel flow}.

\subsection{WALE model}
As baseline method we use the WALE model. The WALE model is an eddy-viscosity model based on the square of the velocity gradient tensor \cite{nicoud_subgrid-scale_nodate}. It has two advantages over the Smagorinsky model in a channel flow simulation. Firstly, it takes into account both strain and rotation rates. And more importantly, its computed eddy-viscosity goes to zero near the channel walls. For incompressible flow, the model is given by
\begin{align}
    \mathbf{m}(\mathbf{v}, C_w) &= \nabla \cdot (2 \nu_t S_{ij}), \\
    \nu_t &=  C_w^2 \Delta^2 \frac{(\mathfrak{S}^d_{ij} \mathfrak{S}^d_{ij})^{3/2}}{(S_{ij} S_{ij})^{5/2}+(\mathfrak{S}^d_{ij} \mathfrak{S}^d_{ij})^{5/4}},
\end{align}
where the symmetric part of the square of the velocity gradient tensor is given by
\begin{equation}
    \mathfrak{S}^d_{ij} = \frac{1}{2}(g_{ik}g_{kj}+g_{jk}g_{ki}).
\end{equation}
Here, $g_{ij} = \partial v_i / \partial x_j$ is the velocity gradient tensor, $S_{ij} = 1/2(\partial v_i/ \partial x_j + \partial v_j/ \partial x_i)$ is the strain rate tensor, $\Delta$ is the filter width (again set to the LF grid size), and $C_w$ is the WALE constant.

\subsection{Fitting eddy-viscosity model constants}
\begin{figure}
    \centering
    \includegraphics[width=0.7\linewidth]{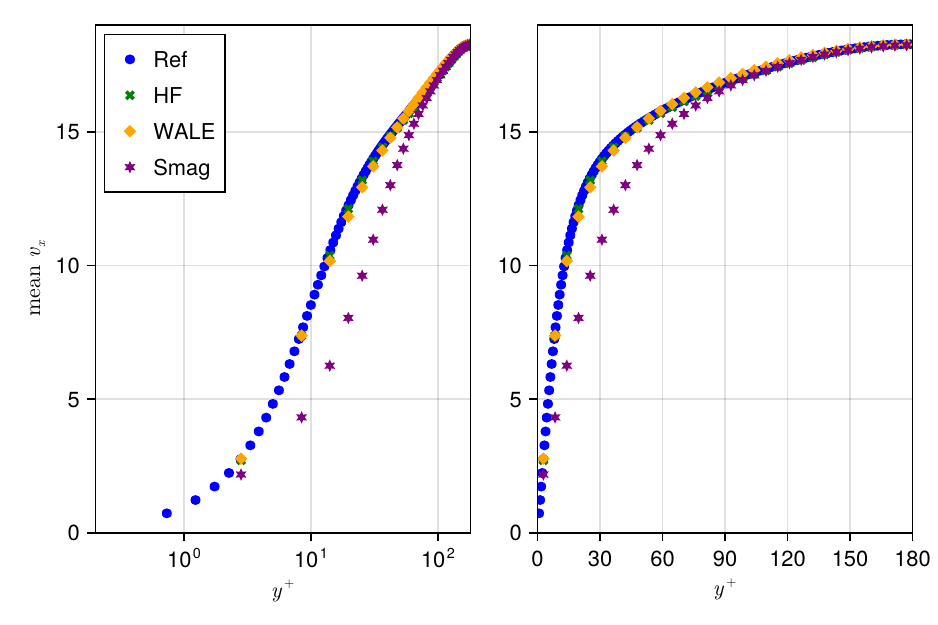}
    \caption{Mean x-velocity profiles over 10 time units of tuned eddy-viscosity models compared to coarse-grained HF solution and reference.}
    \label{fig:x-velocity profiles tuned eddyvisc models}
\end{figure}

To calibrate the WALE and Smagorinsky models, we adjust their constants so that the mean streamwise velocity at the channel center matches that of the high-fidelity simulation over the first 10 time units. Table \ref{tab:tuned eddy viscosity constants} summarizes the tuned values. Figure \ref{fig:x-velocity profiles tuned eddyvisc models} compares the resulting mean x-velocity profiles, plotted against wall distance in both linear and logarithmic scales. Here we show the profile obtained from DNS by \cite{vreman_comparison_2014}, the mean profile from our 10 time unit HF simulation and the profiles of the calibrated Smagorinsky and WALE simulations.

\begin{table}[H]
    \centering
    \begin{tabular}{c|c|c|c}
        & HF & Smag & WALE \\
        &  & $C_s = 0.13$ & $C_w = 0.53$\\ \hline
        mean $v_x(y=1)$ & 18.267 & 18.300 & 18.265
    \end{tabular}
    \caption{Calibrated eddy-viscosity model constants.}
    \label{tab:tuned eddy viscosity constants}
\end{table}

\subsection{TO model}
The TO LRS method was trained on the same 10 time units of the high-fidelity simulation. First, we ran an LF simulation which tracked the reference trajectories of the six QoIs, obtaining a series of subgrid scale corrections. Then we trained an LRS model to predict these subgrid scale corrections, see Section \ref{section: Time-Series Models for SGS Correction}. We evaluated models with history lengths of 5 and 10, which previously performed well on the HIT test case.

On this test case these models lead to instable simulations if there is no regularization. However, introducing a small regularization ($\lambda = 10^{-4}$) was sufficient to get stable simulations.

\subsection{Results}
All subgrid scale models were evaluated in LF simulations over 100 time units. For the TO LRS models we again ran five replica simulations.

\begin{figure}
    \centering
    \includegraphics[width=0.7\linewidth]{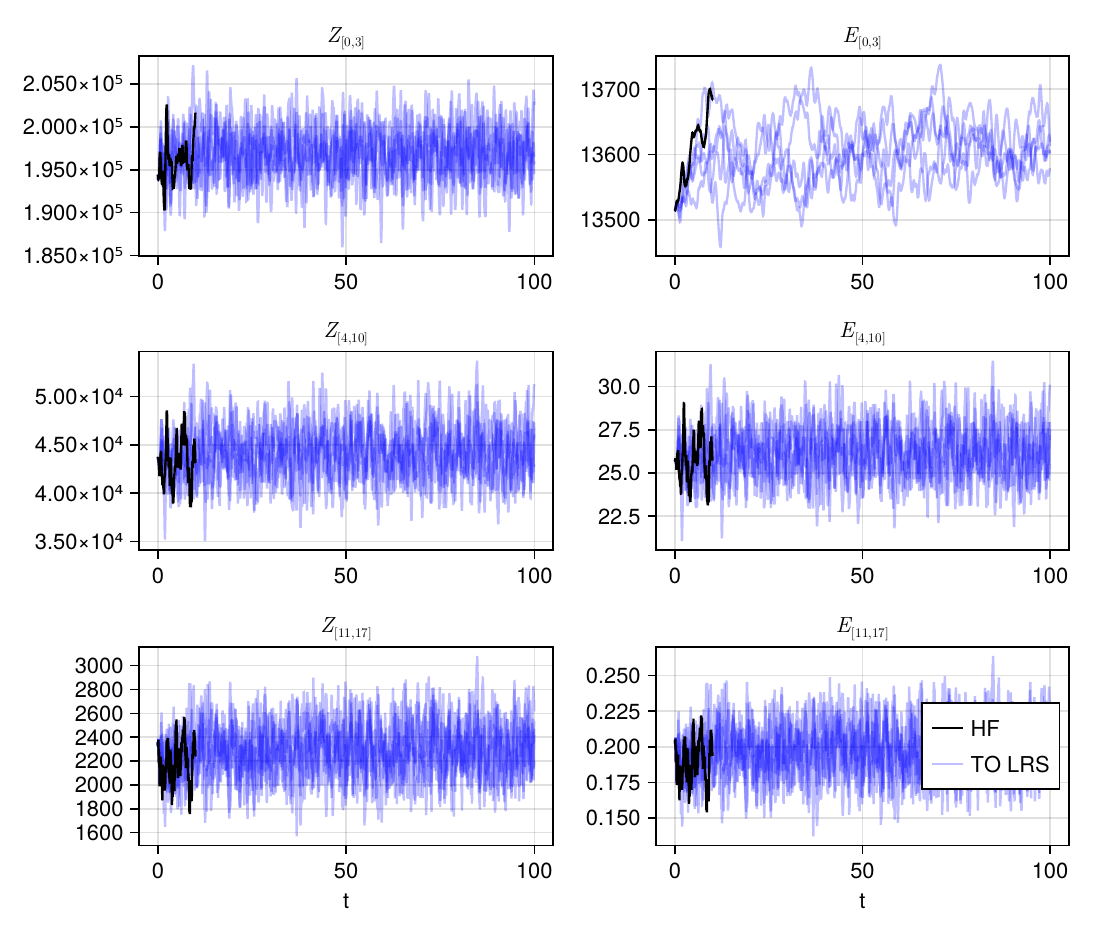}
    \caption{QoI trajectories for TO LRS model with
history length 5 and for the HF training data.}
    \label{fig:channel qoi trajectories TO LRS5}
\end{figure}

\begin{figure}
    \centering
    \includegraphics[width=0.7\linewidth]{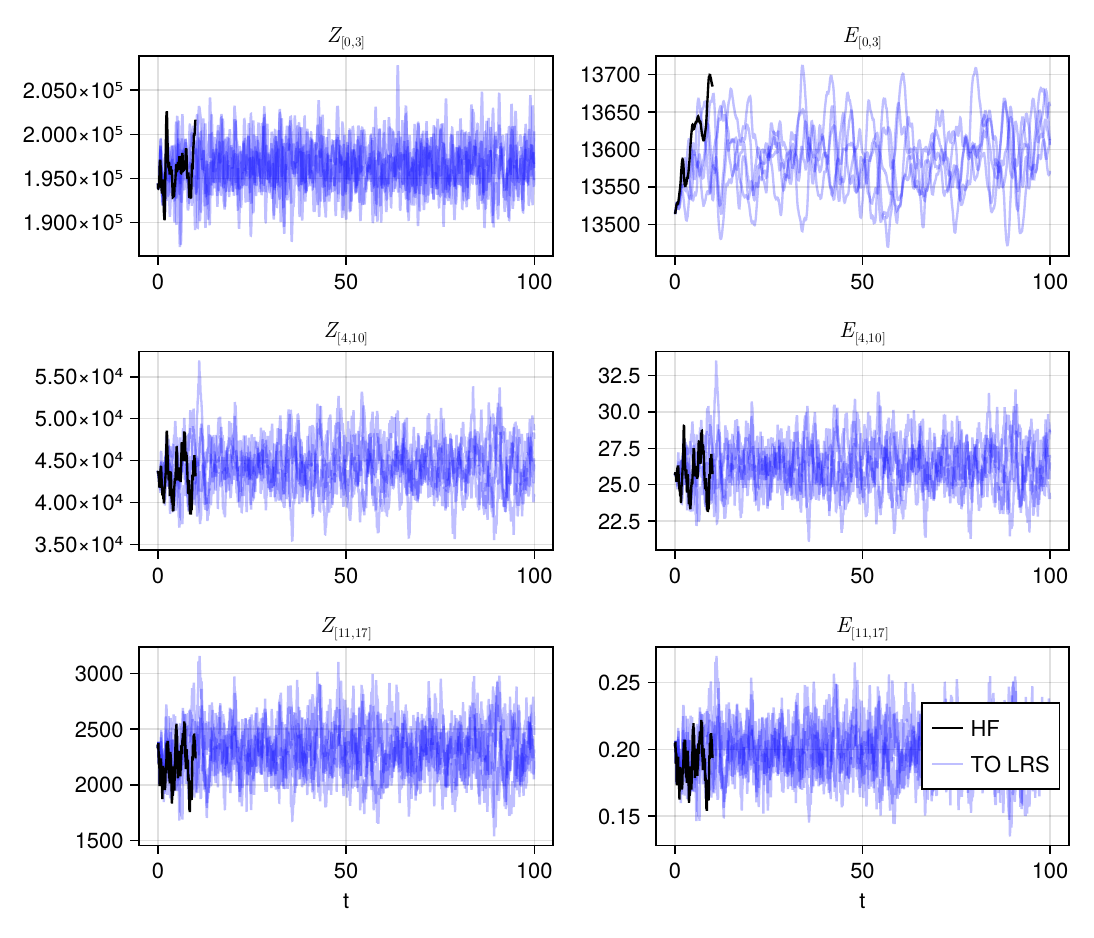}
    \caption{QoI trajectories for TO LRS model with
history length 10 and for the HF training data.}
    \label{fig:channel qoi trajectories TO LRS10}
\end{figure}

Figures \ref{fig:channel qoi trajectories TO LRS5} and \ref{fig:channel qoi trajectories TO LRS10} show the trajectories of the six scale-aware QoIs for TO LRS with history lengths of 5 and 10, respectively. These plots include the corresponding trajectories from the HF simulation, which served as the training data. Across all five replicas, the TO LRS models produced stable simulations, and the QoI trajectories remained close to the range of values seen during training. This indicates that the TO LRS models generalize consistently over longer time horizons, without changing the distribution of the QoIs much.

\begin{figure}
    \centering
    \includegraphics[width=0.7\linewidth]{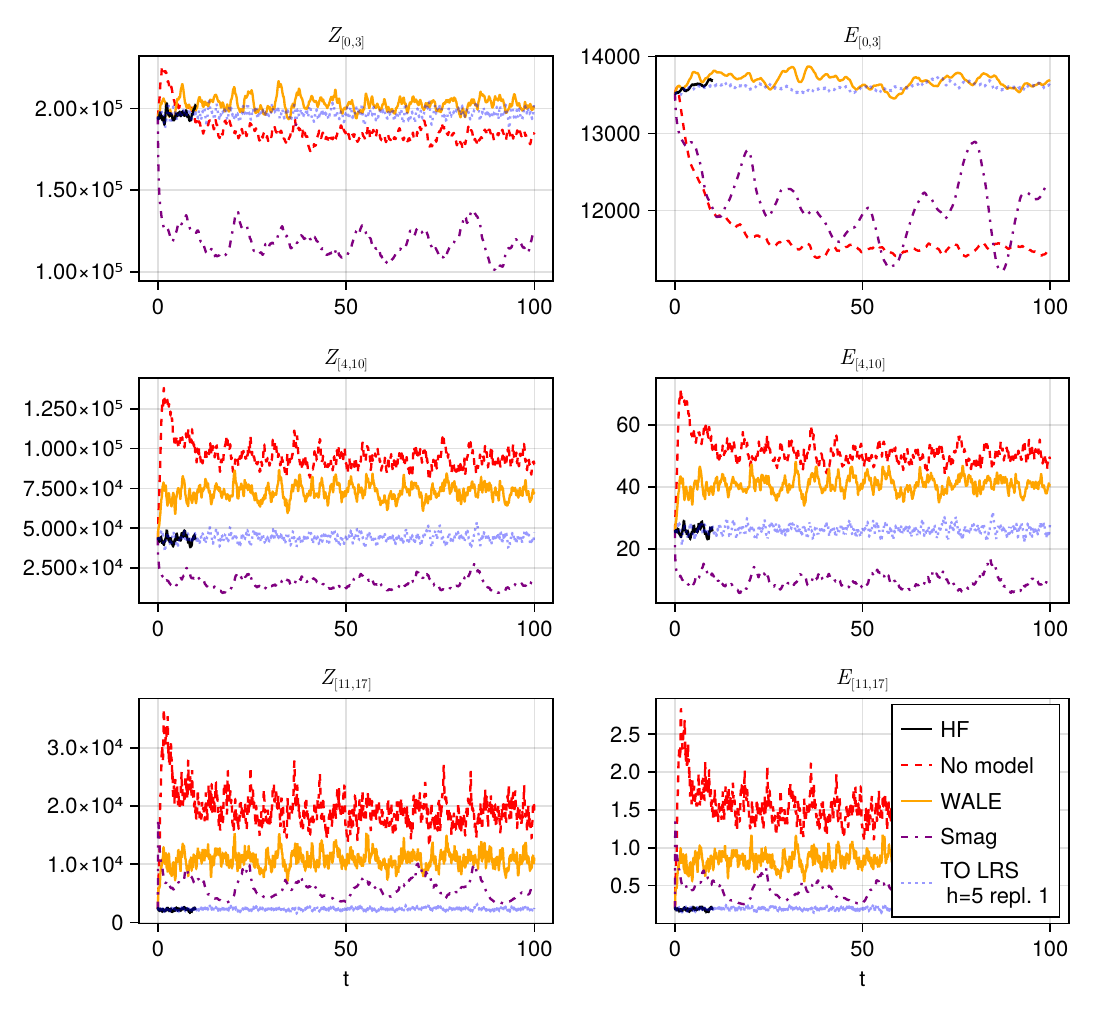}
    \caption{QoI trajectories for eddy-viscosity models, for the HF training data, and the first replica of the TO LRS model with history length 5.}
    \label{fig:channel qoi trajectories eddyvisc models}
\end{figure}

In contrast, Figure \ref{fig:channel qoi trajectories eddyvisc models} presents the QoI trajectories for simulations using the eddy-viscosity models (Smagorinsky and WALE) as well as a simulation without SGS model. The Smagorinsky model and the no model simulations both exhibit significant loss of energy in the largest scales. additionally, the build-up of energy in the smaller scales is apparent for the simulation without SGS model. Interestingly, a similar small-scale energy accumulation is observed for the WALE model. This leads to trajectories with long-term distributions that clearly diverge from the HF data. However, the WALE model performs well when looking at the large-scale energy. 

\begin{figure}
    \centering
    \includegraphics[width=0.7\linewidth]{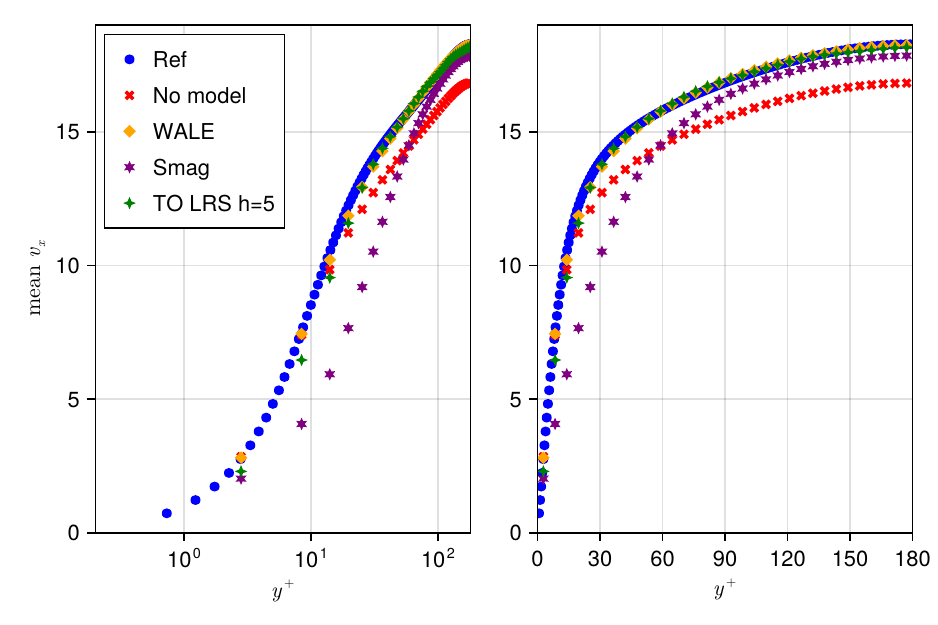}
    \caption{Mean x-velocity profiles over 100 time units in LF simulations with various subgrid scale terms compared to reference.}
    \label{fig:x-velocity profiles online simulations}
\end{figure}

To further assess model performance, Figure \ref{fig:x-velocity profiles online simulations} shows the mean streamwise velocity profile across the channel for various SGS models. The profile from the first TO LRS replica with history length 5 is shown, though other replicas yielded nearly indistinguishable profiles. Among all models, WALE achieves the best match to the reference profile, particularly near the wall. The TO LRS model also performs well, accurately capturing the velocity profile away from the wall, even though it was not explicitly trained to do so. However, it leads to a to low streamwise velocity in the 4 grid points closest to the wall. In contrast, the Smagorinsky model fails to reproduce the correct profile, reflecting its known difficulty in handling wall-bounded turbulence.

\begin{figure}
    \begin{subfigure}[b]{0.49\textwidth}
        \centering
        \includegraphics[width = \linewidth]{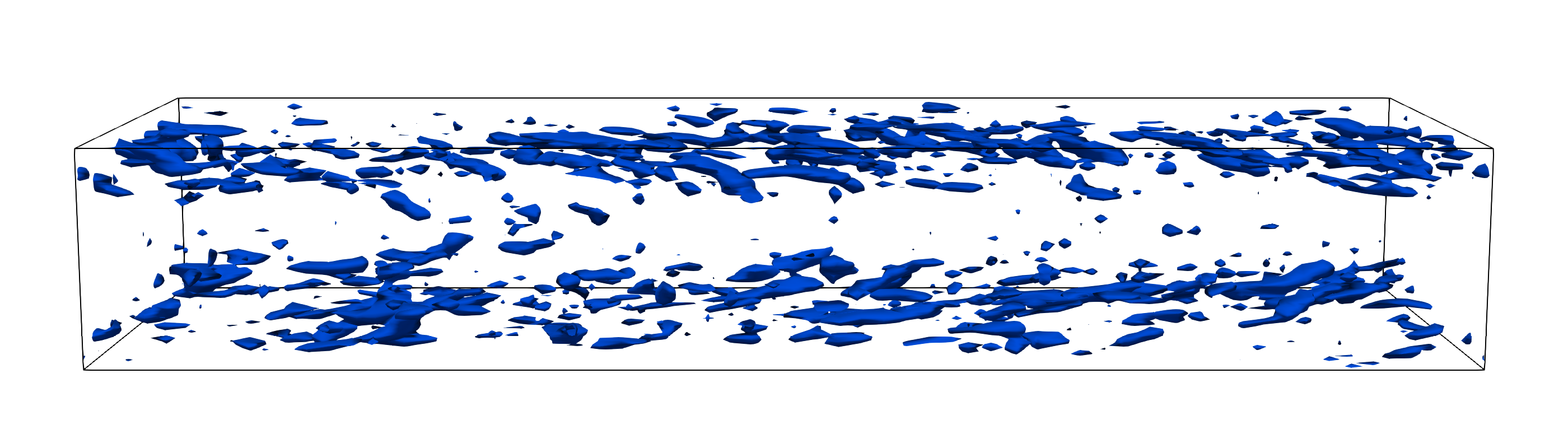}
        \caption{Filtered HF}
    \end{subfigure}
    \begin{subfigure}[b]{0.49\textwidth}
        \centering
        \includegraphics[width = \linewidth]{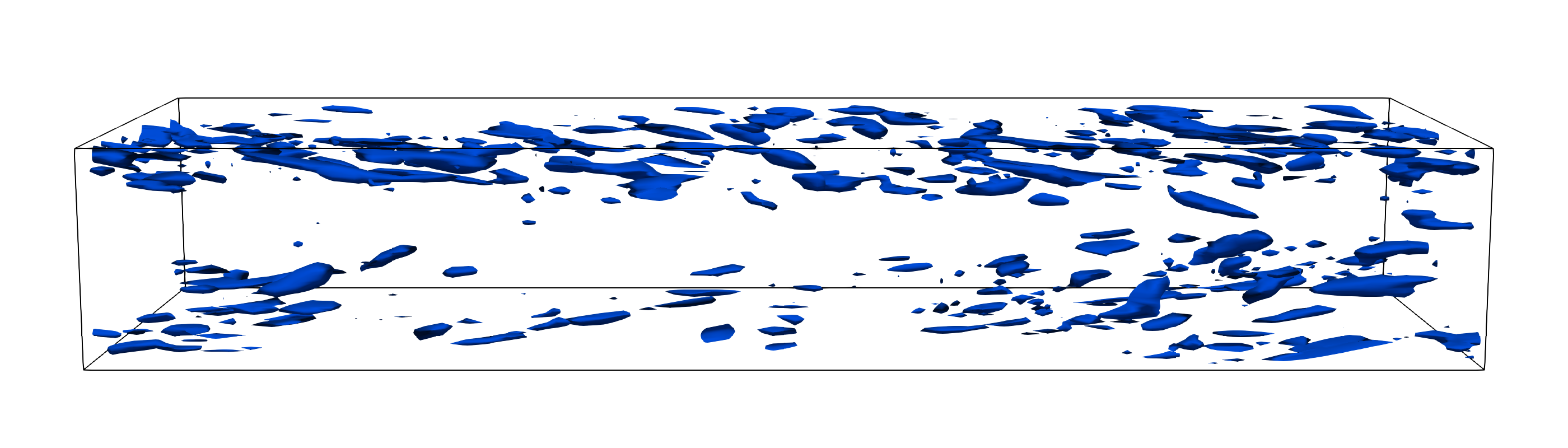}
        \caption{TO LRS h=5}
    \end{subfigure}
    \begin{subfigure}[b]{0.49\textwidth}
        \centering
        \includegraphics[width = \linewidth]{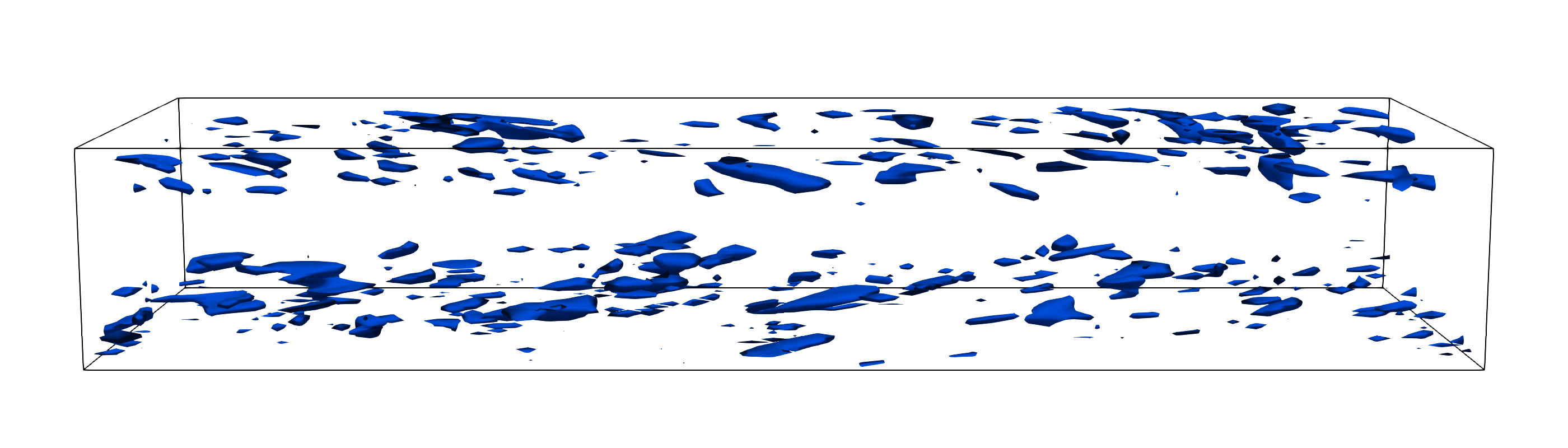}
        \caption{WALE}
    \end{subfigure}
    \begin{subfigure}[b]{0.49\textwidth}
        \centering
        \includegraphics[width = \linewidth]{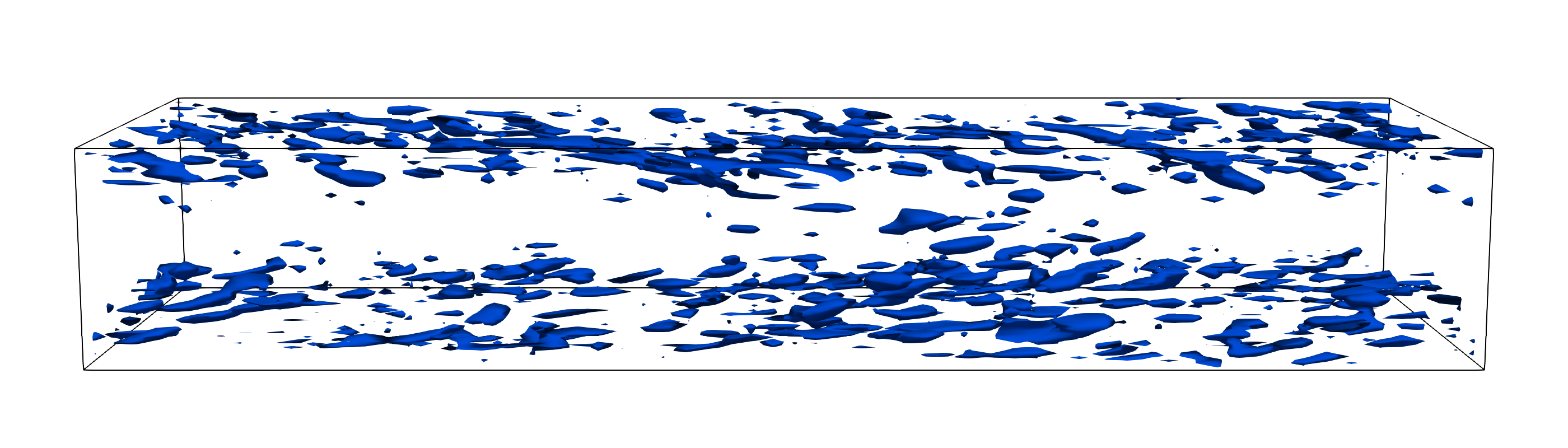}
        \caption{No model}
    \end{subfigure}
    \caption{Turbulent vortices in long term simulations at $T=100$ compared to the filtered HF solution at $T=10$, visualized via isocontours $Q = 100$.}
    \label{fig:Turbulent structures channel}
\end{figure}

Finally, Figure \ref{fig:Turbulent structures channel} visualizes the turbulent structures at the end of the LF simulations using isocontours of the second invariant of the velocity gradient tensor. These are compared against the filtered HF solution at $T=10$. The difference between the LF simulations is less pronounced than in the HIT test case. The LF simulation without SGS term again shows more numerous vortices. But in this test case the filtered HF solution also has more vortices than the LF simulations with SGS terms. Still, a notable result is that the stochastic corrections introduced by the TO LRS method do not break the coherent vortex structures.


\section{Conclusion} \label{section 6: conclusion}
We have shown how the tau-orthogonal method can be extended to three-dimensional flows, a challenge for many data-driven approaches due to cubic scaling of the number of (unclosed) degrees of freedom that need to be modeled. As in the original two-dimensional method, we therefore reformulate the problem of modeling the high-dimensional SGS term as a closure problem for a small set of scale-aware QoIs: energy and enstrophy in large, mid, and small scales. Moreover, we extended the method to be QoI-state dependent, including history. This was achieved by combining a regularized least squares fit with a multivariate Gaussian distribution fitted to the residual (the LRS model).
This new time-series prediction model can easily be fitted to training data obtained from tracking reference QoI trajectories, since it involves a modest number of parameters (in the order of 100 - 1000).

We conducted a hyperparameter study on three-dimensional homogeneous isotropic turbulence, from which we concluded that the LRS model exhibits robustness to the choice of history length, with a broad range of near-optimal values. We also saw that the LRS model outperformed the Smagorinsky model in terms of obtaining correct long-term distributions for the QoIs. Furthermore, it improved on our previously introduced data-driven noise model, which had no dependence on QoI state. 

Remarkably, despite being trained only on reproducing QoI trajectories, the TO LRS model appears to capture other salient features of the turbulent flow. Specifically, it accurately reproduces the averaged kinetic energy spectrum and preserves coherent turbulent structures in the simulated fields, on par with both the Smagorinsky model and coarse-grained high-fidelity solutions.

To test the applicability of this approach to heterogeneous flows, we applied the TO LRS model to a turbulent channel flow. Despite being agnostic to geometry and wall-normal distance, the TO LRS model produced stable simulations with QoI trajectories that remained consistent with the training data. In contrast, the WALE model led to a slight energy build-up in the small-scale QoIs. The TO LRS model captured the mean velocity profile in the channel well, except near the wall, where its performance degraded slightly.

In summary, the TO LRS method provides a simple, data-driven approach to modeling SGS effects that generalizes in time beyond the training domain. This makes it a promising alternative to classical SGS models and traditional deep learning approaches for a broad range of flow configurations. Main avenues for future work will involve improving the basis functions in the TO method for better near-wall behavior and extending the method to non-structured grids.

\section*{Software \& data availability}

The implementation of the methods which are presented in this paper is publicly available in the repository ``TO\_IncompressibleNavierStokes.jl" \cite{TO_Incompressiblenavierstokesjl_2025}. This repository contains scripts and data to reproduce all plots in this paper.

\section*{Funding}

This research is funded by the Netherlands Organization for Scientific Research (NWO) through the ENW-M1 project ``Learning small closure models for large multiscale problems” (OCENW.M.21.053). This work used the Dutch national e-infrastructure with the support of the SURF Cooperative using grant no.\ EINF-9277.

\section*{CRediT author statement}

{\bf Rik Hoekstra}: Conceptualization, Data curation, Formal analysis, Investigation, Methodology, Resources, Software, Validation, Visualization, original draft, review \& editing.
{\bf Wouter Edeling}: Conceptualization; Formal analysis; Funding acquisition; Investigation; Methodology; Project administration;   Supervision; Validation; review \& editing.

\section*{Declaration of competing interest}

The authors declare that they have no known competing financial interests or personal relationships that could have appeared to influence the work reported in this paper.

\section*{Declaration of Generative AI and AI-assisted technologies in the writing process}

During the preparation of this work the authors used ChatGPT in order to rewrite sections of the manuscript. After using this tool/service, the authors reviewed and edited the content as needed and take full responsibility for the content of the published article.

\bibliographystyle{plainnat}
\bibliography{refs}

\begin{thebibliography}{35}
\providecommand{\natexlab}[1]{#1}
\providecommand{\url}[1]{\texttt{#1}}
\expandafter\ifx\csname urlstyle\endcsname\relax
  \providecommand{\doi}[1]{doi: #1}\else
  \providecommand{\doi}{doi: \begingroup \urlstyle{rm}\Url}\fi

\bibitem[Agdestein and Sanderse(2025)]{agdestein_discretize_2024}
S.D. Agdestein and B.~Sanderse.
\newblock Discretize first, filter next: Learning divergence-consistent closure models for large-eddy simulation.
\newblock \emph{Journal of Computational Physics}, 522:\penalty0 113577, 2025.
\newblock \doi{https://doi.org/10.1016/j.jcp.2024.113577}.

\bibitem[Agdestein et~al.(2025)Agdestein, Ciarella, Sanderse, and Hoekstra]{agdesteinincompressiblenavierstokesjl_2024}
S.D. Agdestein, S.~Ciarella, B.~Sanderse, and R.~Hoekstra.
\newblock \emph{IncompressibleNavierStokes.jl (v3.0.0) [Software]}.
\newblock Zenodo, April 2025.
\newblock \doi{https://doi.org/10.5281/zenodo.15294602}.

\bibitem[Chouippe and Uhlmann(2015)]{chouippe_forcing_2015}
A.~Chouippe and M.~Uhlmann.
\newblock Forcing homogeneous turbulence in direct numerical simulation of particulate flow with interface resolution and gravity.
\newblock \emph{Physics of Fluids}, 27\penalty0 (12), 2015.
\newblock \doi{https://doi.org/10.1063/1.4936274}.

\bibitem[Edeling and Crommelin(2020)]{edeling_reducing_2020}
W.N. Edeling and D.T. Crommelin.
\newblock Reducing data-driven dynamical subgrid scale models by physical constraints.
\newblock \emph{Computers \& Fluids}, 201:\penalty0 104470, 2020.
\newblock \doi{https://doi.org/10.1016/j.compfluid.2020.104470}.

\bibitem[Ephrati(2025)]{ephrati2024probabilistic}
S.R. Ephrati.
\newblock Probabilistic data-driven turbulence closure modeling by assimilating statistics.
\newblock \emph{Journal of Computational Physics}, page 114234, 2025.
\newblock \doi{https://doi.org/10.1016/j.jcp.2025.114234}.

\bibitem[Eswaran and Pope(1988)]{eswaran_examination_1988}
V.~Eswaran and S.B. Pope.
\newblock An examination of forcing in direct numerical simulations of turbulence.
\newblock \emph{Computers \& Fluids}, 16\penalty0 (3):\penalty0 257--278, 1988.
\newblock \doi{https://doi.org/10.1016/0045-7930(88)90013-8}.

\bibitem[Germano et~al.(1991)Germano, Piomelli, Moin, and Cabot]{germano1991dynamic}
M.~Germano, U.~Piomelli, P.~Moin, and W.H. Cabot.
\newblock A dynamic subgrid-scale eddy viscosity model.
\newblock \emph{Physics of Fluids A: Fluid Dynamics}, 3\penalty0 (7):\penalty0 1760--1765, 1991.
\newblock \doi{https://doi.org/10.1063/1.857955}.

\bibitem[Gimenez et~al.(2025)Gimenez, S{\'\i}vori, Larreteguy, Monta{\~n}o, Aguerre, Nigro, and Idelsohn]{gimenez_multiscale_2025}
J.M. Gimenez, F.M. S{\'\i}vori, A.E. Larreteguy, S.I. Monta{\~n}o, H.J. Aguerre, N.M. Nigro, and S.R. Idelsohn.
\newblock A multiscale {Pseudo}-{DNS} approach for solving turbulent boundary-layer problems.
\newblock \emph{Computer Methods in Applied Mechanics and Engineering}, 437:\penalty0 117804, 2025.
\newblock \doi{https://doi.org/10.1016/j.cma.2025.117804}.

\bibitem[Girimaji(2024)]{girimaji_turbulence_2024}
S.S. Girimaji.
\newblock Turbulence closure modeling with machine learning: A foundational physics perspective.
\newblock \emph{New Journal of Physics}, 26\penalty0 (7):\penalty0 071201, 2024.
\newblock \doi{10.1088/1367-2630/ad6689}.

\bibitem[Gottwald et~al.(2017)Gottwald, Crommelin, and Franzke]{franzke_stochastic_2016}
G.A. Gottwald, D.T. Crommelin, and C.L.E. Franzke.
\newblock Stochastic climate theory.
\newblock In C.L.E. Franzke and T.J. O’Kane, editors, \emph{Nonlinear and Stochastic Climate Dynamics}, page 209–240. Cambridge University Press, 2017.
\newblock \doi{https://doi.org/10.1017/9781316339251.009}.

\bibitem[Guan et~al.(2023)Guan, Subel, Chattopadhyay, and Hassanzadeh]{guan_learning_2023}
Y.~Guan, A.~Subel, A.~Chattopadhyay, and P.~Hassanzadeh.
\newblock {Learning physics-constrained subgrid-scale closures in the small-data regime for stable and accurate LES}.
\newblock \emph{Physica D: Nonlinear Phenomena}, 443:\penalty0 133568, 2023.
\newblock \doi{https://doi.org/10.1016/j.physd.2022.133568}.

\bibitem[Guillaumin and Zanna(2021)]{guillaumin_stochastic-deep_2021}
A.P. Guillaumin and L.~Zanna.
\newblock Stochastic-deep learning parameterization of ocean momentum forcing.
\newblock \emph{Journal of Advances in Modeling Earth Systems}, 13\penalty0 (9):\penalty0 e2021MS002534, 2021.
\newblock \doi{https://doi.org/10.1029/2021MS002534}.

\bibitem[Hoekstra et~al.(2024)Hoekstra, Crommelin, and Edeling]{hoekstra2024Reduced_data-driven}
R.~Hoekstra, D.T. Crommelin, and W.N. Edeling.
\newblock Reduced data-driven turbulence closure for capturing long-term statistics.
\newblock \emph{Computers \& Fluids}, 285:\penalty0 106469, 2024.
\newblock ISSN 0045-7930.
\newblock \doi{https://doi.org/10.1016/j.compfluid.2024.106469}.

\bibitem[Hoekstra et~al.(2025)Hoekstra, Agdestein, Ciarella, and Sanderse]{TO_Incompressiblenavierstokesjl_2025}
R.~Hoekstra, S.D. Agdestein, S.~Ciarella, and B.~Sanderse.
\newblock \emph{{TO\_IncompressibleNavierStokes.jl} (v0.1) [Software]}.
\newblock Zenodo, July 2025.
\newblock \doi{https://doi.org/10.5281/zenodo.15861632}.

\bibitem[Kurz et~al.(2023)Kurz, Offenhäuser, and Beck]{kurz_Deep_reinforcement_2023}
M.~Kurz, P.~Offenhäuser, and A.~Beck.
\newblock Deep reinforcement learning for turbulence modeling in large eddy simulations.
\newblock \emph{International Journal of Heat and Fluid Flow}, 99:\penalty0 109094, 2023.
\newblock \doi{https://doi.org/10.1016/j.ijheatfluidflow.2022.109094}.

\bibitem[Kurz et~al.(2025)Kurz, Beck, and Sanderse]{kurz_harnessing_2025}
M.~Kurz, A.~Beck, and B.~Sanderse.
\newblock Harnessing equivariance: Modeling turbulence with graph neural networks.
\newblock \emph{arXiv preprint arXiv:2504.07741}, 2025.
\newblock \doi{https://arxiv.org/abs/2504.07741}.

\bibitem[Langford and Moser(1999)]{langford1999IDEALLES}
J.A. Langford and R.D. Moser.
\newblock Optimal {LES} formulations for isotropic turbulence.
\newblock \emph{Journal of fluid mechanics}, 398:\penalty0 321--346, 1999.
\newblock \doi{https://doi.org/10.1017/S0022112099006369}.

\bibitem[Lesieur et~al.(2005)Lesieur, M{\'e}tais, and Comte]{lesieur_large-eddy_2005}
M.~Lesieur, O.~M{\'e}tais, and P.~Comte.
\newblock \emph{Large-eddy simulations of turbulence}.
\newblock Cambridge university press, 2005.
\newblock \doi{https://doi.org/10.1017/CBO9780511755507}.

\bibitem[Li et~al.(2022)Li, Peng, Yuan, and Wang]{li_fourier_2022}
Z.~Li, W.~Peng, Z.~Yuan, and J.~Wang.
\newblock Fourier neural operator approach to large eddy simulation of three-dimensional turbulence.
\newblock \emph{Theoretical and Applied Mechanics Letters}, 12\penalty0 (6):\penalty0 100389, 2022.
\newblock \doi{https://doi.org/10.1016/j.taml.2022.100389}.

\bibitem[Ling and Lozano-Duran(2025)]{ling_numerically_2025}
Y.~Ling and A.~Lozano-Duran.
\newblock Numerically consistent data-driven subgrid-scale model via data assimilation and machine learning.
\newblock In \emph{AIAA SCITECH 2025 Forum}, page 1280, 2025.
\newblock \doi{https://doi.org/10.2514/6.2025-1280}.

\bibitem[List et~al.(2022)List, Chen, and Thuerey]{list2022learned}
B.~List, L.~Chen, and N.~Thuerey.
\newblock Learned turbulence modelling with differentiable fluid solvers: physics-based loss functions and optimisation horizons.
\newblock \emph{Journal of Fluid Mechanics}, 949:\penalty0 A25, 2022.
\newblock \doi{https://doi.org/10.1017/jfm.2022.738}.

\bibitem[List et~al.(2025)List, Chen, Bali, and Thuerey]{list_differentiability_2025}
B.~List, L.~Chen, K.~Bali, and N.~Thuerey.
\newblock Differentiability in unrolled training of neural physics simulators on transient dynamics.
\newblock \emph{Computer Methods in Applied Mechanics and Engineering}, 433:\penalty0 117441, 2025.
\newblock \doi{https://doi.org/10.1016/j.cma.2024.117441}.

\bibitem[Liu et~al.(2022)Liu, Yu, Huang, Liu, and Lu]{liu_investigation_2022}
B.~Liu, H.~Yu, H.~Huang, N.~Liu, and X.~Lu.
\newblock Investigation of nonlocal data-driven methods for subgrid-scale stress modeling in large eddy simulation.
\newblock \emph{AIP Advances}, 12\penalty0 (6), 2022.
\newblock \doi{https://doi.org/10.1063/5.0094316}.

\bibitem[Moin and Kim(1980)]{moin_numerical_1980}
P.~Moin and J.~Kim.
\newblock On the numerical solution of time-dependent viscous incompressible fluid flows involving solid boundaries.
\newblock \emph{Journal of computational physics}, 35\penalty0 (3):\penalty0 381--392, 1980.
\newblock \doi{https://doi.org/10.1016/0021-9991(80)90076-5}.

\bibitem[Moser et~al.(1999)Moser, Kim, and Mansour]{moser_direct_1999}
R.D. Moser, J.~Kim, and N.N. Mansour.
\newblock Direct numerical simulation of turbulent channel flow up to {$Re_{\tau}$}=590.
\newblock \emph{Phys. fluids}, 11\penalty0 (4):\penalty0 943--945, 1999.
\newblock \doi{https://doi.org/10.1063/1.869966}.

\bibitem[Nicoud and Ducros(1999)]{nicoud_subgrid-scale_nodate}
F.~Nicoud and F.~Ducros.
\newblock Subgrid-scale stress modelling based on the square of the velocity gradient tensor.
\newblock \emph{Flow, turbulence and Combustion}, 62\penalty0 (3):\penalty0 183--200, 1999.
\newblock \doi{https://doi.org/10.1023/A:1009995426001}.

\bibitem[Park and Choi(2021)]{park_toward_2021}
J.~Park and H.~Choi.
\newblock Toward neural-network-based large eddy simulation: Application to turbulent channel flow.
\newblock \emph{Journal of Fluid Mechanics}, 914:\penalty0 A16, 2021.
\newblock \doi{https://doi.org/10.1017/jfm.2020.931}.

\bibitem[Perezhogin et~al.(2023)Perezhogin, Zanna, and Fernandez-Granda]{perezhogin_generative_2023}
P.~Perezhogin, L.~Zanna, and C.~Fernandez-Granda.
\newblock Generative data-driven approaches for stochastic subgrid parameterizations in an idealized ocean model.
\newblock \emph{Journal of Advances in Modeling Earth Systems}, 15\penalty0 (10):\penalty0 e2023MS003681, 2023.
\newblock \doi{https://doi.org/10.1029/2023MS003681}.

\bibitem[Rasp(2020)]{rasp_coupled_2020}
S.~Rasp.
\newblock {Coupled online learning as a way to tackle instabilities and biases in neural network parameterizations: General algorithms and Lorenz 96 case study (v1. 0)}.
\newblock \emph{Geoscientific Model Development}, 13\penalty0 (5):\penalty0 2185--2196, 2020.
\newblock \doi{https://doi.org/10.5194/gmd-13-2185-2020}.

\bibitem[Sanderse et~al.(2025)Sanderse, Stinis, Maulik, and Ahmed]{sanderse_scientific_2024}
B.~Sanderse, P.~Stinis, R.~Maulik, and S.E. Ahmed.
\newblock Scientific machine learning for closure models in multiscale problems: A review.
\newblock \emph{Foundations of Data Science}, 7\penalty0 (1):\penalty0 298--337, 2025.
\newblock \doi{https://doi.org/10.3934/fods.2024043}.

\bibitem[Saura and Gomez(2022)]{saura_subgrid_2022}
N.~Saura and T.~Gomez.
\newblock Subgrid stress tensor prediction in homogeneous isotropic turbulence using {3D}-convolutional neural networks.
\newblock \emph{Preprint at SSRN 4184202}, 2022.
\newblock \doi{https://doi.org/10.2139/ssrn.4184202}.

\bibitem[Shebalin and Woodruff(1997)]{shebalin_kolmogorov_1997}
J.V. Shebalin and S.L. Woodruff.
\newblock Kolmogorov flow in three dimensions.
\newblock \emph{Physics of Fluids}, 9\penalty0 (1):\penalty0 164--170, 1997.
\newblock \doi{https://doi.org/10.1063/1.869159}.

\bibitem[Smagorinsky(1963)]{smagorinsky_general_1963}
J.~Smagorinsky.
\newblock General circulation experiments with the primitive equations: I. the basic experiment.
\newblock \emph{Monthly weather review}, 91\penalty0 (3):\penalty0 99--164, 1963.

\bibitem[Van~Gastelen et~al.(2024)Van~Gastelen, Edeling, and Sanderse]{van_gastelen_energy-conserving_2023}
T.~Van~Gastelen, W.N. Edeling, and B.~Sanderse.
\newblock Energy-conserving neural network for turbulence closure modeling.
\newblock \emph{Journal of Computational Physics}, 508:\penalty0 113003, 2024.
\newblock \doi{https://doi.org/10.1016/j.jcp.2024.113003}.

\bibitem[Vreman and Kuerten(2014)]{vreman_comparison_2014}
A.W. Vreman and J.G.M. Kuerten.
\newblock Comparison of direct numerical simulation databases of turbulent channel flow at {$Re_{\tau}$}=180.
\newblock \emph{Physics of Fluids}, 26\penalty0 (1), 2014.
\newblock \doi{https://doi.org/10.1063/1.4861064}.

\end{thebibliography}

\appendix
\section{Quadrature in the Fourier domain} \label{app: Fourier quadature}
Our scale-aware QoIs are based on local energy/enstrophy density in Fourier space. In this section we show that these densities are related to the total energy in physical space via quadrature.
We rely on the Discrete Fourier transform of the velocity field, which gives Fourier coefficients $\hat{\mathbf{u}}_\mathbf{k}$ such that
\begin{equation}
    \mathbf{u}(x,y,z) \approx \frac{1}{N_x N_y N_z} 
    \sum_{k_1 = 0}^{N_x-1} \sum_{k_2 = 0}^{N_y-1} \sum_{k_3 = 0}^{N_z-1}
    \exp \left[2 \pi i \left(\frac{xk_1}{L_xN_x}+\frac{yk_2}{L_yN_y}+\frac{zk_3}{L_zN_z}\right)\right] \hat{\mathbf{u}}_\mathbf{k},
\end{equation}
where $N_x$ denotes the number of grid points in the $x$-direction and $L_x$ denotes the length of the computational domain in this direction.

Let us now consider the total kinetic energy of a real-valued velocity field $\mathbf{u}$
\begin{align}
    E &=  \frac{1}{2} \int_\Omega {\mathbf{u} \cdot \mathbf{u}} \, \D \boldsymbol{x} \nonumber,\\
    &\approx \frac{1}{2} \int_\Omega \frac{1}{(N_x N_y N_z)^2} 
    \sum_{\mathbf{k}} \sum_{\mathbf{l}}
    \exp \left[2 \pi i \left(\frac{x_1(k_1+l_1)}{L_xN_x}+\frac{x_2(k_2+l_2)}{L_yN_y}+\frac{x_3(k_3+l_3)}{L_zN_z}\right)\right] \hat{\mathbf{u}}_\mathbf{k}\cdot 
    \hat{\mathbf{u}}_\mathbf{l} \, \D \boldsymbol{x} \nonumber,\\
    & = \frac{1}{2}  \frac{\lvert \Omega \rvert}{(N_x N_y N_z)^2} \sum_{\mathbf{k}} \hat{\mathbf{u}}_\mathbf{k}\cdot 
    \hat{\mathbf{u}}_\mathbf{-k}\nonumber,\\
    & = \frac{1}{2}  \frac{\lvert \Omega \rvert}{(N_x N_y N_z)^2} \sum_{\mathbf{k}} \hat{\mathbf{u}}_\mathbf{k}\cdot 
    \text{conj}(\hat{\mathbf{u}}_\mathbf{k}),
\end{align}
where $\lvert \Omega \rvert$ denotes the volume of the computational domain.
Here, we used the orthogonality of Fourier modes to get from the second to the third expression. The last step only holds for fields that are real-valued in physical space.

\section{Regularization in the LRS model}\label{app: regularization}
\begin{figure}
    \centering
    \includegraphics[width=0.7\linewidth]{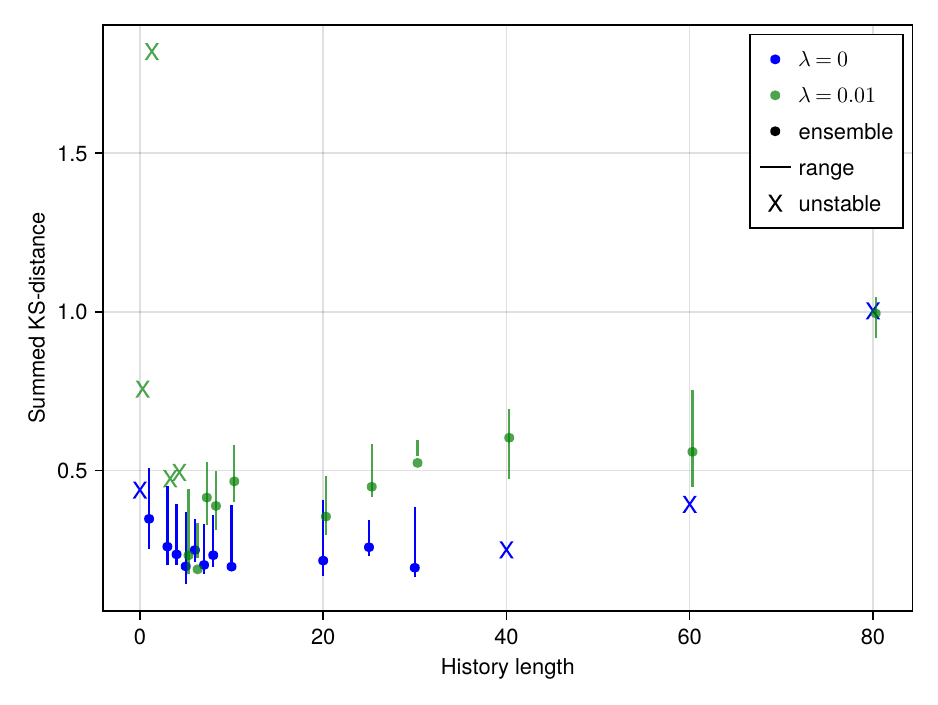}
    \caption{Influence of regularization on predictive quality LRS model in the HIT test case, extrapolating from a 10 time unit training domain to 100 time units. The plotted ranges are based on 5 replica simulations. The ensemble KS-distance results from aggregating the QoI data of all ensemble members. ``X'' markers indicate at least one unstable replica, their location is based on the ensemble KS-distance of the stable part of the trajectories.}
    \label{fig:app Influence of regularization}
\end{figure}

We examine the effect of L2-regularization on the TO LRS models in the HIT test case. Figure~\ref{fig:app Influence of regularization} shows the outcome of applying a mild regularization term ($\lambda = 0.01$). The results indicate that regularization stabilizes models with longer history lengths. Conversely, it causes instability in models with shorter history.

\begin{figure}
    \begin{subfigure}[b]{0.49\textwidth}
        \centering
        \includegraphics[width = \linewidth]{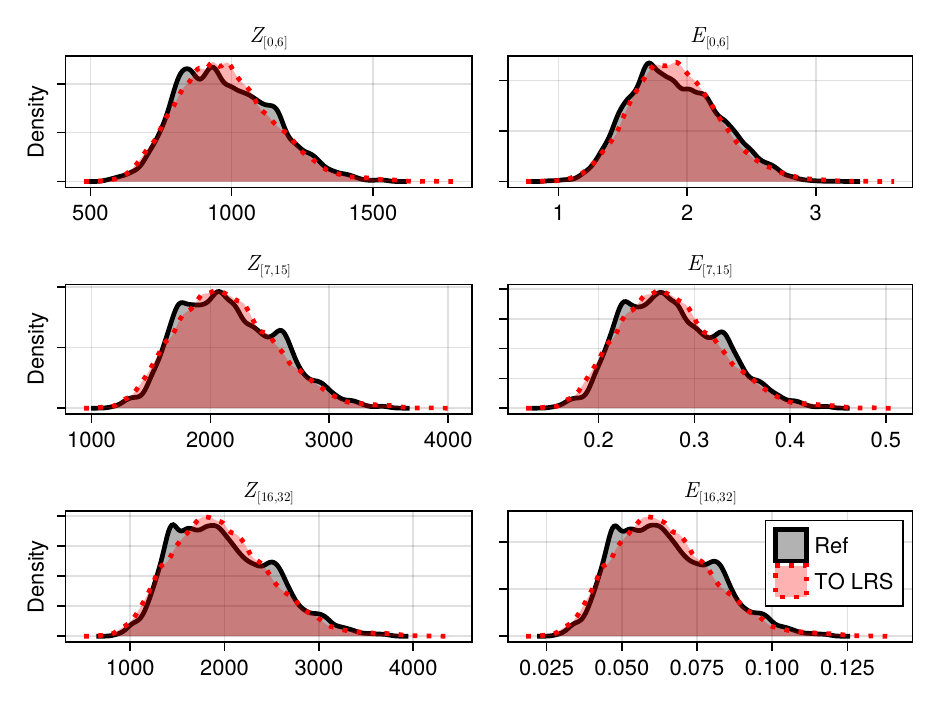}
        \caption{$\lambda = 0$}
    \end{subfigure}
    \begin{subfigure}[b]{0.49\textwidth}
        \centering
        \includegraphics[width = \linewidth]{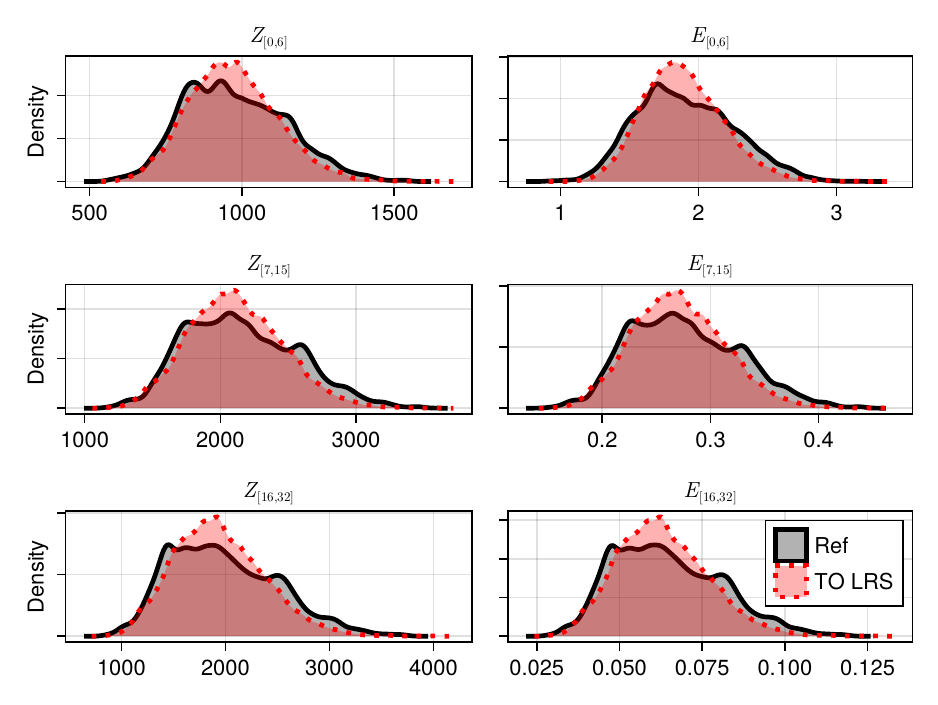}
        \caption{$\lambda = 0.01$}
    \end{subfigure}
    \caption{Long-term QoI distributions in HIT of 5-member ensemble TO LRS with history length 10, with and without regularization.}
    \label{fig: app long-term distr with regularization}
\end{figure}

Overall, the addition of regularization tends to degrade model performance. This is further illustrated in Figure~\ref{fig: app long-term distr with regularization}, which compares the long-term QoI distributions for the regularized and non-regularized TO LRS model with a history length of 10. Regularization in this case results in a to low variance.

\section{Scale-aware filters applied to initial field channel flow} \label{app: Scale-aware filters applied to initial field channel flow}
\begin{figure}
\centering
    \begin{subfigure}[b]{0.49\textwidth}
        \centering
        \caption{$\lVert R_{[0,3]}\bar{\boldsymbol{\omega}}\rVert$}
        \includegraphics[width = \linewidth]{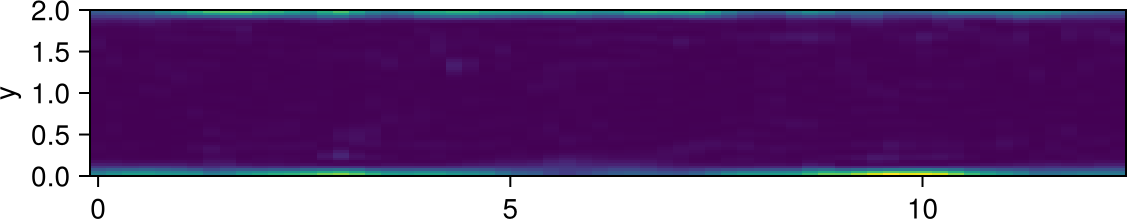}
    \end{subfigure}
    \begin{subfigure}[b]{0.49\textwidth}
        \centering
        \caption{$\lVert R_{[0,3]}\bar{\mathbf{u}}\rVert$}
        \includegraphics[width = \linewidth]{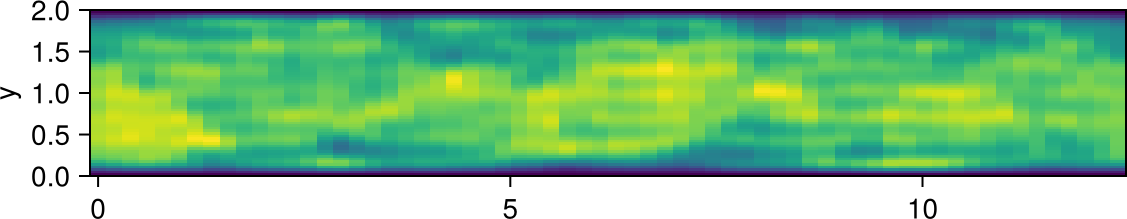}
    \end{subfigure}
    \begin{subfigure}[b]{0.49\textwidth}
        \centering
        \caption{$\lVert R_{[4,10]}\bar{\boldsymbol{\omega}}\rVert$}
        \includegraphics[width = \linewidth]{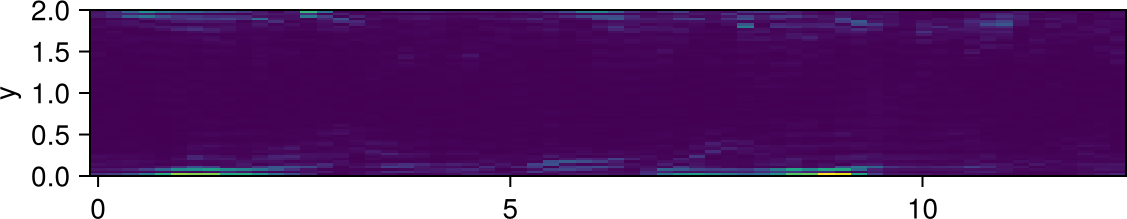}
    \end{subfigure}
    \begin{subfigure}[b]{0.49\textwidth}
        \centering
        \caption{$\lVert R_{[4,10]}\bar{\mathbf{u}}\rVert$}
        \includegraphics[width = \linewidth]{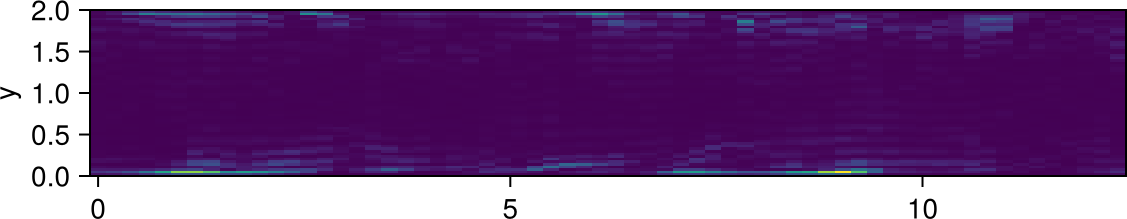}
    \end{subfigure}
    \begin{subfigure}[b]{0.49\textwidth}
        \centering
        \caption{$\lVert R_{[11,17]}\bar{\boldsymbol{\omega}}\rVert$}
        \includegraphics[width = \linewidth]{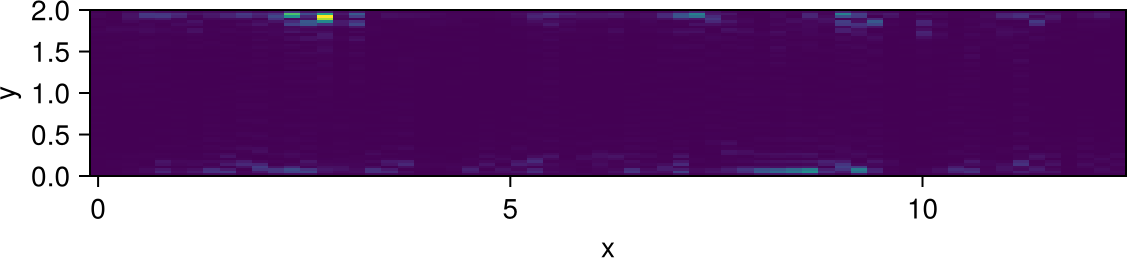}
    \end{subfigure}
    \begin{subfigure}[b]{0.49\textwidth}
        \centering
        \caption{$\lVert R_{[11,17]}\bar{\mathbf{u}}\rVert$}
        \includegraphics[width = \linewidth]{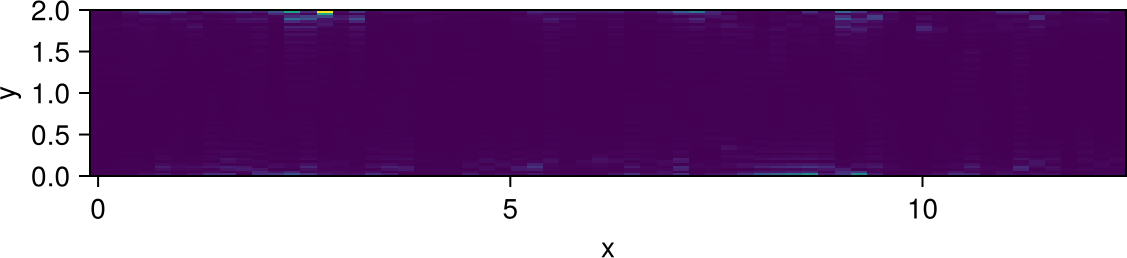}
    \end{subfigure}
    \caption{Scale-aware filters applied to initial turbulent field of channel flow test case.}
    \label{fig: Scale-aware filtered channel.}
\end{figure}

For our channel flow test case we defined six scale-aware QoIs: energy and enstrophy in three wavenumber bands: $[0, 3]$, $[4, 10]$, and $[11, 17]$. To get a feeling for these QoIs, we plotted the velocity and vorticity fields which result from filtering. Figure \ref{fig: Scale-aware filtered channel.} shows the velocity and vorticity magnitudes of the scale-aware filtered initial field for: $R_{[l,m]}\mathbf{u}$ and $R_{[l,m]}\boldsymbol{\omega}$. We see that the large-scale energy is located in the bulk of the channel. The other QoIs are highest near the walls, as expected. 

\end{document}